\documentclass[structabstract]{aa}  

\usepackage{graphicx}
\usepackage{natbib}
\usepackage{txfonts}
\usepackage{bbold}   

\begin{document}

\title{A two-column formalism for time-dependent modelling of stellar convection}
\subtitle{I. Description of the method}

\author{Alexander St\"okl}

\institute{CRAL, Universit\'e de Lyon, CNRS (UMR5574), \'Ecole Normale Sup\'erieure de Lyon, F-69007 Lyon, France \\
           \email{alexander.stoekl@ens-lyon.fr}	}

\date{Received ???; accepted ???}

\abstract
{In spite of all the advances in multi-dimensional hydrodynamics, 
investigations of stellar evolution and stellar pulsations still depend on one-dimensional computations. 
This paper devises an alternative to the mixing-length theory or turbulence models usually adopted 
in modelling convective transport in such studies.}
{The present work attempts to develop a time-dependent description of convection, which reflects 
the essential physics of convection and that is 
only moderately dependent on numerical parameters 
and far less time consuming than existing multi-dimensional hydrodynamics computations.}
{Assuming that the most extensive convective patterns generate the majority of convective transport, 
the convective velocity field is described using two parallel, radial columns to represent up- and downstream flows. 
Horizontal exchange, in the form of fluid flow and radiation, over their connecting interface couples the two columns 
and allows a simple circulating motion.
The main parameters of this convective description have straightforward geometrical meanings, 
namely the diameter of the columns (corresponding to the size of the convective cells) 
and the ratio of the cross-section between up- and downdrafts.
For this geometrical setup, the time-dependent solution of the equations of radiation hydrodynamics is computed from 
an implicit scheme that has the advantage of being unaffected by the Courant-Friedrichs-Lewy time-step limit.
This implementation is part of the 
TAPIR-Code (short for {\bf T}he {\bf a}da{\bf p}tive, {\bf i}mplicit {\bf R}HD-Code).}
{To demonstrate the approach, results for convection zones in Cepheids are presented.
The convective energy transport and convective velocities agree with expectations for Cepheids and the scheme 
reproduces both the kinetic energy flux and convective overshoot. 
A study of the parameter influence shows that the type of solution derived for these stars is in fact fairly robust
with respect to the constitutive numerical parameters.
}
{}


\maketitle

\section{Introduction}

Convection is one of the persistent problems in stellar astrophysics.
Almost all stars contain regions where convective transport is important; 
in the photosphere, in the envelope, or in the interior where nuclear burning occurs.
A description of convection is therefore an essential ingredient to all types
of investigations of stellar structure and evolution.
Unfortunately, due to its nonlinear, nonlocal, and multi-length-scale nature, 
modelling convection turns out to be an intricate problem.

Conceptually, there are several
different approaches to the numerical simulation of convective transport in stars.
The most straightforward approach is the time-dependent solution of the equations of radiation hydrodynamics 
in a 3D (or at least 2D) domain to compute the convective flow patterns directly.
However, this process is very expensive in computing time, prohibitively so for some applications. 
This particularly applies to problems with a large difference in relevant timescales,  
for instance between the thermal and acoustic timescale in the stellar interior, 
or the hydrodynamics and radiative timescale in the outer layers of luminous stars. 
An additional limit is imposed by the restricted spatial resolution. 
Even the most elaborate, high-resolution simulations of stellar convection
are only capable of resolving the largest scales in the convective velocity field.
The effect of turbulence on smaller length scales is effectively ignored, 
even though it is a possible interpretation 
to attribute the intrinsic numerical dissipation of the scheme to some (unknown) unresolved turbulence.
In particular, hydrodynamics codes often include artificial viscosity for numerical stability, and
`unresolved turbulence' is the only physical mechanism that could be used to justify 
the inclusion of this additional dissipation.

Despite the poor description of turbulence in hydrodynamics computations,
multi-dimensional simulations of solar granulation \citep[e.g.][]{SN98,Asplund2000,Wedemeyer2004} 
achieve remarkable quantitative agreement with observations.


A completely different and more subtle approach than trying to resolve turbulence on large numerical grids
are convection models that use an equation, or a set of equations, 
to describe convective transport either with a heuristic parametrization or based on turbulence theory.

The most widely used of these 1D-descriptions is the well known mixing-length theory (MLT) \citep{MLT,CoxandGiuli}, 
which originates in ideas of \citet{Prandtl1925}.
The MLT has been remarkably successful in stellar astrophysics in application to stars from white dwarfs to super giants, 
probably because of its simple yet flexible parametrization 
and its robust reference to the adiabatic temperature gradient. 


Modern alternatives to the MLT are convection models in which 
the original single-eddy assumption of the MLT has been replaced with a full spectrum of turbulence 
\citep[e.g.][]{CM91,C96,CGM96} by either assuming or computing a turbulent energy spectrum.
For some types of stars, these models can also avoid the mixing-length scale as a free parameter.
Convection models of this type are included in many stellar evolution codes as an alternative to MLT.

However, all of these models are, in a similar way to the original MLT, 
local theories that do not provide any information about overshoot.
In stellar-evolution codes, this deficit is overcome by adding
overshoot by means of a separate (typically diffusive) parametrization.


Another type of one-equation models consists of a 
time-dependent equation for the turbulent kinetic energy \citep[e.g.][]{stell,kuhfuss,GM92}
using heuristic approximations for individual terms. 
Containing a diffusion term for the turbulent kinetic energy, they also allow a simple form of non-locality.
These convection models are mainly geared towards 
computations of stellar pulsations \citep[e.g.][]{BS1994, FeuchtingerptII, KBSC2002}
where one is interested in the time-dependence of the convective transport. 


Finally, the most complete way to model turbulence is the `Reynolds stress approach' by 
solving a set of moment equations \citep{canuto1992, canuto1993, canuto1997, Xiong1989, Xiongetal1997}
that are terminated by a closure model at the third or fourth order.
These closures are based on either the quasi-normal approximation for the fourth-order moments 
or use a parametrization with reference to 
measured data, hydrodynamics (large-eddy) simulations, or concepts such as the `plume model' (see below). 
Turbulence models
of this type are able to describe convective transport in a time-dependent and non-local way.
For applications of the Reynolds stress model to stellar surface convection zones, see 
\citet{Kupka99,KuMo2002,MoKu2004}.


The two-column scheme, presented in this paper, is in-between the above categories,
combining a hydrodynamics simulation of the convective fluid flow 
with a parametrized, predefined geometry of the flow patterns.
This setup is almost as simple as a 1D description; 
it describes up- and downstream with two parallel radial columns, 
and fluid flow over an interface between those two columns allows a basic circulating convective motion. 
The two-column model 
could therefore be regarded as a simplistic 2D hydrodynamics scheme, 
limiting the horizontal range of the grid to just two cells.
The very coarse description of the convective velocity field 
effectively implies that the most extensive convective flow patterns generate the majority of the convective transport.
However, this assumption does not differ significantly 
from what is assumed in multi-dimensional hydrodynamics computations 
that are also unable to resolve the full spectrum of convective turbulence. 
Since multi-dimensional hydrodynamics achieve, in spite of this limitation, a good agreement with observations, 
the scenario of macroscopic convective patterns with distinct up- and downstream regions 
and little sub-structure (as also observed in the solar granulation) 
appears to be a sufficient description of the actual physics \citep{nordlundetal1997}. 
The two-column approach may therefore also give reasonable results.

In the two-column scheme, the convective flux is computed directly from hydrodynamics 
and not from a heuristic model. 
Although it is not without numerical parameters, there is no `mixing-length parameter', nor anything equivalent. 
The method is also intrinsically non-local and the thickness of convective regions and the amount of overshoot  
are obtained consistently.

The basic idea of modelling convection using separate radial stratifications 
for up- and downstream regions is in fact an old one.
In the 1960's to 1970's, predating advances in computing power that enabled 2D and 3D hydrodynamics computations to become possible, 
several similar two- or multi-stream models were devised.
From observations, the solar granulation pattern appeared to be separable to almost distinct hot and cool areas; 
it was therefore a logical first step to place two stratifications next to each other 
in order to construct more realistic models of the solar photosphere \citep{MG1969, nordlund76}. 
More recently a two-stream model was applied by \cite{lesaffre2005} to investigate 
the convective Urca process in supernova-progenitor white dwarfs.
In geophysics, a similar concept, known as `plume model', was introduced by \citet{MTT1956}. 
In its basic form, this model considers only plumes that are immersed in a static, surrounding medium, 
although there are also models that consider both up- and downdrafts and their interaction 
\citep[e.g.][]{Telford1970, WA1986, CB1987, RSM1992}.
The idea of separated up- and downwards streams was also used to construct closures for turbulence models 
\citep{AMF1997, ZGLM1999, LR2001, GH2002, GHRS2005, CCH2007}


However, all existing multi-stream models differ from the present attempt in that the 
`two-column-scheme' introduced below is based on 
fully implicit time-dependent radiation hydrodynamics in \emph{both} radial and horizontal directions  
without any ad-hoc assumptions or parametrizations for the physical coupling of the two columns.
The term `two-column' (in contrast to `two-stream') was chosen intentionally
because the discretization scheme resembles that of two 1D discretizations placed beside each other in two parallel columns.
In analogy to `2D' for `two-dimensional', we will occasionally refer to `two-column' as `2C' in the following sections.

The remaining paper is structured as follows. 
The next section, Sect.~\ref{sect.2C-scheme}, introduces the two-column discretization scheme and its geometric derivation.  
The equations of radiation hydrodynamics are given in Sect.~\ref{sect.RHDeq-anal} in analytical form,                        
while Sect.~\ref{sect.RHDeq-discr} describes their discretization and considers details                                      
such as artificial viscosity and radiative transport.
Section~\ref{sect.method} presents the deployed solution algorithm,                                                          
followed by a demonstrating example and some parameter studies in Sect.~\ref{sect.results}.                                  
Finally, Sect.~\ref{sect.concl} draws the conclusions and summarizes the paper.                                              
A verification of the method by comparison with detailed 2D hydrodynamics computations as well as 
applications in time-dependent calculations of Cepheids' pulsations 
will be given in the forthcoming part II paper of this series.

\section{The two-column discretization scheme}
\label{sect.2C-scheme}

\begin{figure}[thbp]
\begin{center}
\includegraphics[width=0.7\linewidth, angle=00]{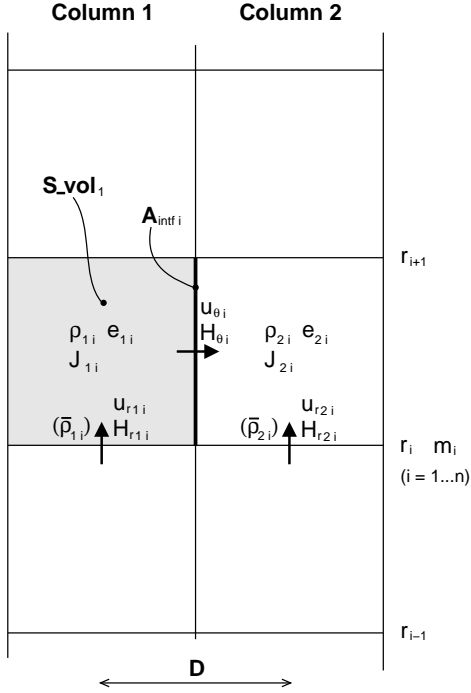}
\end{center}
\caption{
The two-column discretization scheme and the localization of the primary variables (see Table~\ref{tab.primvar}).
$D$ is the typical distance/diameter of the columns.
The area shaded in gray represents the discretization volume ${\rm S\_vol_1}$ used 
for the scalar variables $\rho_1$, $e_1$, $J_1$, and their respective equations.
Advection occurs, as indicated by arrows, over the radial interface as well as over the interface $A_{\rm intf}$ 
between the two columns.
The vector quantities $u_r$, $H_r$, $u_\theta$, and $H_\theta$ are included in their appropriate staggered-mesh location. 
Note that, although not illustrated in the figure, this all occurs in spherical geometry.}
\label{fig.2C-diskr-plus} 
\end{figure}

Figure~\ref{fig.2C-diskr-plus} shows the setup of the two-column discretization scheme 
and the localization of the primary variables 
(for a complete listing of the primary variables, see Table~\ref{tab.primvar}). 
The columns do not correspond directly to an individual convective cell but should be considered as a representation of
all up- and downstream flows, respectively. 
Correspondingly, the interface between the two columns represents the sum of all contact surfaces between up- and downdrafts.

\begin{figure}[thbp]
\begin{center}
\includegraphics[width=0.85\linewidth, angle=00]{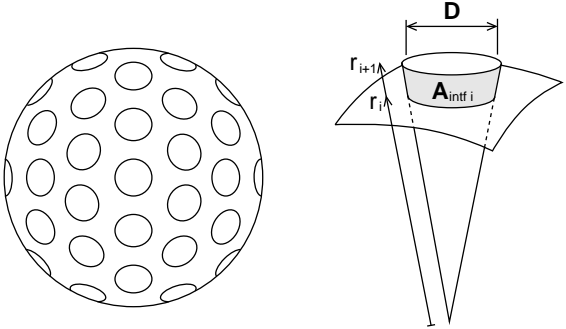}
\end{center}
\caption{Spherical interpretation of the two-column scheme. 
The illustration on the left hand side shows the principle of how $N$ cone-like 
cells are assumed to be distributed over the sphere.
Each of these cells, as sketched on the right, contributes to the interface area $A_{\rm intf}$.
Since the description is symmetric for the two columns (except for the relative cross-sections), 
these cones can be considered to represent either up- or downdrafts.}
\label{fig.Aintf}
\end{figure}

In the 2C-scheme, the horizontal components of fluid flow and radiation, 
which actually occur in both the $\theta$- and $\phi$-direction of spherical geometry, 
are described each by just one `horizontal' variable. 
For these variables, $u_\theta$ and $H_\theta$ respectively (see Table~\ref{tab.primvar}),  
the subscript `$\theta$' does not refer to spherical geometry components 
but is used more generally to denote `horizontal' variables.

The geometrical configuration of the discretization scheme is specified by two parameters. 
The first parameter is the typical horizontal length scale $D$, 
which can be interpreted as the diameter of the convective cells 
or the typical horizontal distance between up- and downdrafts.
In contrast to the sketch in Fig.~\ref{fig.2C-diskr-plus}, the two columns, in general, do not have the same size. 
The second parameter $cf_1$ ($cf$ for \emph{column fraction}) specifies the fraction of the sphere 
that is associated with column 1, and $cf_2$ correspondingly, with $cf_2 = 1 - cf_1$, is that allocated to column 2. 

These two parameters, $D$ and $cf_1$, with their straightforward geometrical meaning 
are the main free parameters of the convection model. 
In Sect.~\ref{sect.horizAdv}, we will introduce a third constitutive parameter correlated with horizontal advection.
Other numerical parameters, such as solution accuracy, radial advection scheme, artificial viscosity, boundary conditions, and grid resolution 
have only a minor effect on the solution.

An equivalent yet more concrete 
quantity than $D$ is the number of convective cells on a sphere $N$. 
In principle, it is possible to assign different values of $N$ to individual shells,  
or to specify an analytical relation that defines $N$, for instance as a function of the radius. 
However, in the absence of a robust physical indication for the behavior of $N$, 
the present implementation uses the same $N$ for every shell independent of the radius. 
That way, the up- and downdrafts are assumed to retain their identity throughout the convective region, 
which appears reasonable for photospheric convection zones of moderate depth.  
The relative cross-sections $cf1$ and $cf2$ must remain constant in all cases since changing their
values would tilt the interface between the columns from the radial (coordinate) direction.

By definition, $D$ and $N$ are related by
\begin{equation}
\label{eq.D}
      D = 2 \, r \, \sqrt{2/N} \; ,
\end{equation}
which is based on the assumption of circular convective cells as sketched in Fig.~\ref{fig.Aintf}.
Since the formalism of the two columns is symmetric apart from in the relative cross-sections,
it makes no difference whether these circular cells are considered as up- or as downdrafts.
The right hand side of Fig.~\ref{fig.Aintf} illustrates the computation of the interface area between the two columns.
Using Eq.~\ref{eq.D} and summing for $N$ columns, we obtain
\begin{equation}
\label{eq.Aintf}
      A_{\rm intf} = \frac{\pi}{2} \sqrt{N/2} \left( r_{i+1}^2 - r_i^2 \right) \; .
\end{equation}
Despite the geometric motivation given in Fig.~\ref{fig.Aintf}, this picture should not be interpreted literally.
By combining both the $\theta$- and $\phi$-directions of spherical geometry
to one generic horizontal variability, the direct correlation with the 
three-dimensional configuration disappears. 
It is therefore not sensible to interpret expressions from the two-column formalism using, 
for example, a specific slice through the setup shown in Fig.~\ref{fig.Aintf}.

However, the only points where the geometrical configuration enters the scheme are the definitions of the 
interface area in Eq.~\ref{eq.Aintf} and of the horizontal derivatives in 
Eqs.~\ref{eq.derivtheta1}~\&~\ref{eq.derivtheta2}.
Since these equations are coupled with each other 
by the requirement to ensure that Gau\ss 's theorem applies also in the discrete case, 
a different geometrical picture would change Eqs.~\ref{eq.Aintf}~--~\ref{eq.derivtheta2} by only a constant factor.

\subsubsection*{Horizontal derivatives}

Using the typical horizontal length scale $D$, 
we can approximate derivatives in the horizontal direction by
\begin{equation}
\label{eq.derivtheta1}
\frac{1}{r} \frac{\partial}{\partial \theta} X \simeq \frac{1}{D} \, \Delta_\theta(X) = \frac{1}{2 r} \, \sqrt{N/2} \; \Delta_\theta(X), 
\end{equation}
where `$\theta$' is again used to denote the generic horizontal extension of the 2C-scheme.

For second-order derivatives with respect to $\theta$, we adopt the estimate
\begin{equation}
\label{eq.derivtheta2}
\frac{1}{r^2} \frac{\partial^2 X}{\partial \theta^2} \simeq \frac{2}{D^2} \, \Delta_\theta(X) 
            = \left\{ \begin{array}{cl}
                    \displaystyle{ \frac{N}{4 r^2} (X_2 - X_1) }  &\quad \mbox{for column 1} \\[5mm]
                    \displaystyle{ \frac{N}{4 r^2} (X_1 - X_2) }  &\quad \mbox{for column 2} 
                    \end{array} \right. \; ,
\end{equation}
which is based on the assumption of a basically periodic variation of $X$ in $\theta$-direction 
(and consequently of $\frac{\partial X}{\partial \theta}$ as well) due to the alternating 
succession of up- and downdraft columns.
This already requires over-stretching of the geometrical picture 
but since these derivatives exist at a less important point 
(in the $\frac{\partial}{\partial \theta}$ term of the viscous forces, Eqs.~\ref{eq.2C-force-radial}~\&~\ref{eq.2C-force-theta}), 
the approximative evaluation of $\frac{\partial^2 X}{\partial \theta^2}$ is acceptable.

\subsection{Scalar discretization}

The discretization of scalar physical variables and equations uses discretization volumes 
similar to that highlighted in gray in Fig.~\ref{fig.2C-diskr-plus}.
The volume of the scalar cells (distinguished by the prefix `S$\_$' for \emph{scalar}) is computed to be the appropriate 
fraction of the shell between the radii $r_i$ and $r_{i+1}$
\begin{equation}\label{eq.S_vols}
   \begin{array}{lcl}
      {\rm S\_vol_1} &=& cf_1 \; \frac{4 \pi}{3} \, \left( r_{i+1}^3 - r_i^3 \right) \\[3pt]
      {\rm S\_vol_2} &=& cf_2 \; \frac{4 \pi}{3} \, \left( r_{i+1}^3 - r_i^3 \right) \;.
   \end{array}
\end{equation}
The advective fluxes (transported `volume' during a time step) 
for the scalar discretization are indicated by arrows in Fig.~\ref{fig.2C-diskr-plus}.
Radial advection consists of two contributions; 
one from the proper motion of the fluid, and one due to movement of the adaptive grid
\begin{equation}\label{eq.S_fluxrad}
   \begin{array}{lcl}
        {\rm S\_flux_1} &=& cf_1 \left[ 4 \pi \, r_i^2 \, u_{1,i}\, \delta t - \frac{4 \pi}{3} \left( {r_i^{\, new}}^3 - {r_i^{\, old}}^3 \right) \right] \\[5pt]
        {\rm S\_flux_2} &=& cf_2 \left[ 4 \pi \, r_i^2 \, u_{2,i}\, \delta t - \frac{4 \pi}{3} \left( {r_i^{\, new}}^3 - {r_i^{\, old}}^3 \right) \right] 
   \end{array}
\end{equation}
where $\delta t$ is the time step during which the grid adaptivity alters the radius of 
the grid point $i$ from $r_i^{\, old}$ to $r_i^{\, new}$.
Note that the individual radial velocities $u_{1,i}$ and $u_{2,i}$ were used in the two columns.

The horizontal advective flux between the two columns is computed to be 
\begin{equation}\label{eq.S_fluxtheta}
        {\rm S\_flux_\theta} = A_{\rm intf} \;  u_{\theta, i} \, \delta t 
                             = \frac{\pi}{2} \sqrt{N/2} \left( r_{i+1}^2 - r_i^2 \right)  u_{\theta, i} \, \delta t \;.
\end{equation}
For both the horizontal velocity $u_\theta$ and the horizontal advective flux ${\rm S\_flux_\theta}$, 
a positive sign corresponds to a fluid flow from column 1 to column 2 by convention.
The analogous convention is also used for the horizontal radiative flux $H_{\theta}$.

\subsection{Vector discretization -- radial}

\begin{figure}[thbp]
\begin{center}
\includegraphics[width=0.6\linewidth, angle=00]{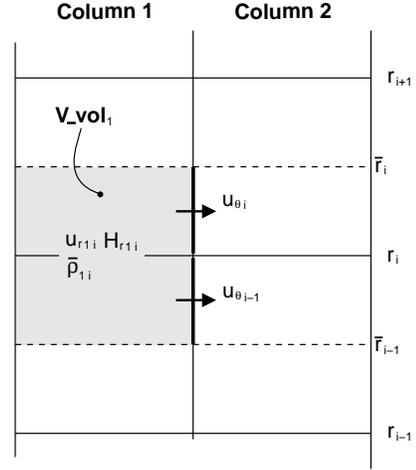}
\end{center}
\caption{Discretization volume for vector variables and equations in radial direction.
The radial boundaries of the cell are defined by the averaged radii $\overline r_i$ and $\overline r_{i-1}$.
The arrows illustrate horizontal advection which is composed of two parts 
correlated with $u_{\theta\,i}$ and $u_{\theta\,i-1}$.}
\label{fig.2C-diskr-Rvector} 
\end{figure}

Vector-type variables and equations are discretized on a staggered mesh where 
the radial part closely resembles that given by \citet{Doetal}.
To define discretization volumes for the radial vector components, 
we start by defining `averaged' radii located 
in between the grid point positions and denoted by $\overline r$
\begin{equation} \label{eq.averageradius}
    \overline r_i^3 \equiv r_{i+\frac{1}{2}}^3 = \frac{1}{2} \, \left( r_i^3 + r_{i+1}^3 \right) \; .
\end{equation}
Figure~\ref{fig.2C-diskr-Rvector} shows two of these averaged radii, $\overline r_i$ and $\overline r_{i-1}$, 
which establish the discretization volumes centered around the grid point $r_i$.
Using the definition of $\overline r$, 
we can now compute the volumes (with the prefix `V$\_$' for \emph{vector}) of these cells
\begin{equation}
   \begin{array}{lcl}
      {\rm V\_vol_1} &=& cf_1 \; \frac{4 \pi}{3} \, \frac{1}{2} \, \left( r_{i+1}^3 - r_{i-1}^3 \right)  \\[3pt]
      {\rm V\_vol_2} &=& cf_2 \; \frac{4 \pi}{3} \, \frac{1}{2} \, \left( r_{i+1}^3 - r_{i-1}^3 \right) \;.
   \end{array}
\end{equation}
For the advective flux in the radial direction, we interpolate the velocity assuming flux conservation, i.e.\ 
\begin{equation}
r_{i+1/2}^2 u_{i+1/2} = \frac{1}{2} \left( r_i^2 u_i + r_{i+1}^2 u_{i+1} \right) \; , 
\end{equation}
and therefore in analogy with Eq.~\ref{eq.S_fluxrad}, we obtain 
\begin{equation}
   \begin{array}{lcrc}
      {\rm V\_flux_1} &=& cf_1 \Bigl[ 4 \pi \, \frac{1}{2} \left( r_i^2 \, u_{1,i} + r_{i+1}^2 \, u_{1,i+1} \right)  \, \delta t      & \\
                      & &             - \frac{4 \pi}{3} \left( {\overline r_i^{\, new}}^3 - {\overline r_i^{\, old}}^3 \right) \Bigr] & \\[7pt]
      {\rm V\_flux_2} &=& cf_2 \Bigl[ 4 \pi \, \frac{1}{2} \left( r_i^2 \, u_{2,i} + r_{i+1}^2 \, u_{2,i+1} \right)  \, \delta t      & \\
                      & &             - \frac{4 \pi}{3} \left( {\overline r_i^{\, new}}^3 - {\overline r_i^{\, old}}^3 \right) \Bigr] & .
   \end{array}
\end{equation}
As indicated by the arrows in Fig.~\ref{fig.2C-diskr-Rvector}, 
the horizontal flux between the two 
radial vector volumes ${\rm V\_vol_1}$ and ${\rm V\_vol_2}$ is composed of two parts
correlated with $u_{\theta\,i}$ and $u_{\theta\,i-1}$
\begin{equation}
      {\rm V\_flux_\theta} = \frac{\pi}{2} \sqrt{N/2} \left[ 
      \left( \overline r_i^2   - r_i^2 \right) u_{\theta,i}   + 
      \left( r_i^2 - \overline r_{i-1}^2 \right) u_{\theta,i-1} \right] \, \delta t \; .
\end{equation}

\subsection{Vector discretization -- horizontal}

\begin{figure}[thbp]
\begin{center}
\includegraphics[width=0.8\linewidth, angle=00]{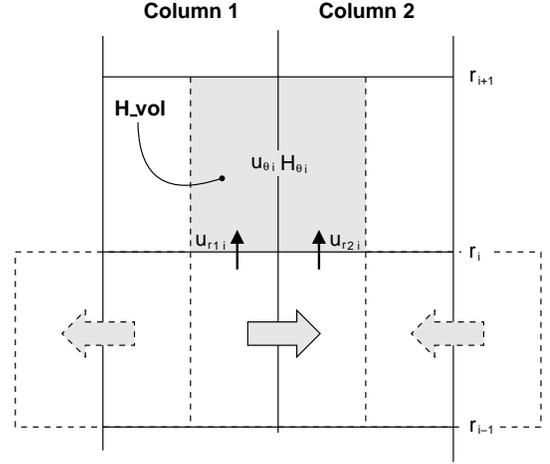}
\end{center}
\caption{Discretization volume for the horizontal components of vector variables and equations.
The radial advective flux is composed of two parts that, except for a factor $1/2$, 
resemble the scalar radial fluxes ${\rm S\_flux_1}$ and ${\rm S\_flux_2}$.
The gray arrows in the lower part illustrate the mirroring principle 
used when considering the two columns as part of a sequence of up- and downdrafts.
In that picture, the two outer cells with the left-hand arrows are actually one and the same.
}
\label{fig.2C-diskr-Tvector} 
\end{figure}

The discretization of the horizontal components of vector variables and equations 
(namely of $u_\theta$ and $H_\theta$) 
uses discretization volumes as illustrated in Fig.~\ref{fig.2C-diskr-Tvector}.
The corresponding volumes and fluxes are labeled with the prefix `H$\_$' for \emph{horizontal}.
The discretization cell with ${\rm H\_vol}$ is centered on 
the interface between the columns and considers half the volume of the sphere 
\begin{equation}
\label{eq.H_vol}
      {\rm H\_vol} =  \frac{1}{2} \; \frac{4 \pi}{3} \, \left( r_{i+1}^3 - r_i^3 \right) \; . 
\end{equation}
The second half of the volume is assumed to mirror the physics of the first one. 
Even though the discretization volume ${\rm H\_vol}$ represents only one half of the shell, 
it therefore describes the horizontal components of the entire sphere.

From comparison with Eq.~\ref{eq.S_vols}, we observe that ${\rm H\_vol} = 1/2 \,\left( {\rm S\_vol_1} + {\rm S\_vol_2} \right)$; 
the flux in the radial direction ${\rm H\_flux}$ is assembled in the same way from 
${\rm S\_flux_1}$ and ${\rm S\_flux_2}$ (see Eq.~\ref{eq.S_fluxrad})
\begin{equation}
\label{eq.H_fluxrad}
   \begin{array}{lcrc}
      {\rm H\_flux}  &=&\frac{1}{2} \; \Bigl[ 4 \pi \, r_i^2 \, \left( cf_1 u_{1,i} + cf_2 u_{2,i} \right) \, \delta t    & \\
                     & &                      - \frac{4 \pi}{3} \left( {r_i^{\, new}}^3 - {r_i^{\, old}}^3 \right) \Bigr] & .
   \end{array}
\end{equation}
Since the discretization scheme represents a large number of convective cells distributed over the sphere, 
the two columns can be considered as part of a sequence of alternating up- and downdrafts. 
Along this sequence, the direction of horizontal fluid flow and radiation switches its sign repeatedly.
This implies that a right-hand orientated flow 
(as illustrated in the lower part of Fig.~\ref{fig.2C-diskr-Tvector}) is confronted with 
an equal flow in the opposing direction when reaching the (in this case) right hand cell boundary.
To allow for this effect, a dissipation term was included in the equation of motion 
that could be interpreted as annihilation of the momentum of the two opposing flows 
\begin{equation} 
F_{\rm anhl} = - \left| {\rm S\_fluxH} \right| \; \left( cf_1 \rho_1 + cf_2 \rho_2 \right) u_\theta
\end{equation}
where $\left( cf_1 \rho_1 + cf_2 \rho_2 \right) u_\theta$ is the momentum in the cell and $\left| {\rm S\_fluxH} \right|$ 
describes the volume fraction swept against the horizontal boundary.
In the equation of internal energy, this dissipated energy enters as 
\begin{equation}
E_{\rm anhl} = \left| {\rm S\_fluxH} \right| \; \left( cf_1 \rho_1 + cf_2 \rho_2 \right) u^2_\theta \; .
\end{equation}
The location at which this energy is deposited 
depends on the direction of horizontal flow. 
In the case sketched in Fig.~\ref{fig.2C-diskr-Tvector}, the dissipated energy would be deposited into column 2.

As we will make use of it 
in the set of discrete equations (Table~\ref{tab.discSetofEq}),  
we finally define $\overline \rho_\theta$, 
the averaged density appropriate for the `horizontal' discretization volume 
\begin{equation}
\label{eq.rhoH_avg}
\overline \rho_\theta = \left( cf_1 \rho_1 + cf_2 \rho_2 \right) \; .
\end{equation}
Despite the similar notation, 
this \emph{horizontally} averaged density should not be mistaken for 
the \emph{radially} averaged densities $\overline \rho_1$ and $\overline \rho_2$
that are introduced in Sect.~\ref{sect.rho_avg}.

\subsection{Horizontal advection and energy conservation}

\subsubsection*{Horizontal advection}
\label{sect.horizAdv} 

\begin{figure}[thbp]
\begin{center}
\includegraphics[width=0.65\linewidth, angle=00]{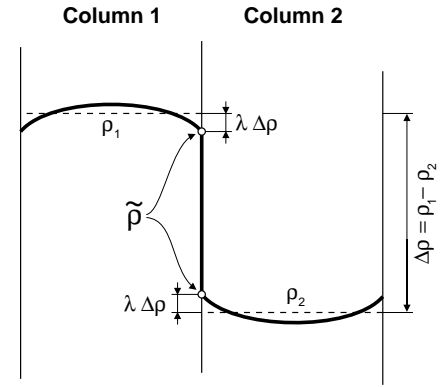}
\end{center}
\caption{
Reconstruction scheme for horizontal advection.
Allowing some variation in the variables (here $\rho$ as an example) 
across the width of the columns
decreases the contrast at the interface between up- and downdraft.
For computation of the horizontally advected quantity $\widetilde{\rho}$, this variation 
is assumed to be proportional to the contrast between the two columns $\Delta \rho$.
The parameter $\lambda$ with $0 \le \lambda \le 1/2$ allows for a continuous transition between 
donor cell and centered advection.}
\label{fig.H-advec} 
\end{figure}
Advection in the radial direction is considered using a second-order van~Leer advection scheme \citep{vL74,vL77}, but
for horizontal advection, i.e.\ fluid flow from one column to the other, 
a higher order scheme is obviously not applicable.
The most straightforward approach is donor cell advection, but, in general, we can allow some variation within the columns, as 
illustrated in Fig.~\ref{fig.H-advec} for the example of density. 
This produces the following advection scheme
\begin{equation}
\label{eq.rho-H-advec}
\widetilde{\rho}   = \left\{ \begin{array}{cl}
	            (1-\lambda) \; \rho_1 + \lambda \; \rho_2 &\quad \mbox{for Col.~1} \rightarrow \mbox{Col.~2} \\[3pt]
	            (1-\lambda) \; \rho_2 + \lambda \; \rho_1 &\quad \mbox{for Col.~1} \leftarrow  \mbox{Col.~2}
                    \end{array} \right. \: ,
\end{equation}
where we adopt the convention of denoting advected quantities with an overhead tilde.
From Eq.~\ref{eq.rho-H-advec}, one recovers both simple donor cell advection for $\lambda = 0$ 
and centering between  $\rho_1$ and $\rho_2$ for $\lambda = 1/2$. 
However, it is advisable to retain a small value of $\lambda$, i.e.\ resulting in an advection scheme similar to donor-cell,  
since centered advection can produce unrealistic values for advected radial momentum: 
as part of a large circulating fluid flow, the convective motions in the two columns correspond to each other.
Centering the momentum between the up- and downdraft column therefore gives almost zero momentum. 
Consequently, the up- and downdraft flows lose hardly any momentum due to horizontal advection, 
even if there is a large horizontal exchange of mass and internal energy. 
This mechanism only affects the momentum as it is the sole advected quantity 
where the values for the two columns usually have opposing signs.

Despite this potentially unphysical behavior for larger values of $\lambda$, 
the formalism in Eq.~\ref{eq.rho-H-advec} provides an additional free parameter 
for adjusting the convection zones obtained from the 2C-scheme with reference to established results.

\subsubsection*{Energy conservation}
\label{sect.enecons} 

The equations of radiation hydrodynamics (Eqs.~\ref{eq.cont_anal}~--~\ref{eq.radflx_anal}) 
are discretized conservatively.
Hence, the scheme conserves mass, momentum, internal energy, as well as the moments of radiation. 
Although analytically equivalent, this does not translate into conservation of the \emph{total} energy in the discrete case.
Usually, this is not crucial for 1D computations. 
Moreover, the adaptive grid provides a fine grid resolution for all gradients 
and accordingly minimizes spatial discretization errors. 
In case of the horizontal components in the 2C-scheme, 
we now have to consider a very coarse spatial representation where, in particular, 
advection from one column to another requires some attention. 

In the present discretization, advection of total energy consists of three components: 
internal, radiative, and kinetic energy.
The former two are treated accurately by the advection terms in the corresponding 
equations of internal energy and radiation energy.
In contrast, the advection of kinetic energy is modeled only indirectly by density and momentum transport. 
Analytically, advection of kinetic energy, momentum, and density are related by 
\begin{equation}
\vec \nabla \cdot \left( \vec u {\textstyle \frac{1}{2}} \rho \vec u^2 \right) = 
	\vec u \cdot                      \vec \nabla \cdot \left( \vec u \, \rho \vec u \right) -
	{\textstyle \frac{1}{2}} \vec u^2 \vec \nabla \cdot \left( \vec u \, \rho        \right) \; .
\end{equation}
Integration over a cell volume provides the discrete (approximate) equivalent 
\begin{equation}
 \sum \limits_i \widetilde{{\textstyle \frac{1}{2}} \rho u^2_i} {\rm Flux}_i \simeq
	 u                           \sum \limits_i \widetilde{\rho u_i} \; {\rm Flux}_i -
	{\textstyle \frac{1}{2}} u^2 \sum \limits_i \widetilde{\rho_i}   \; {\rm Flux}_i
\end{equation}
where the quantities with a tilde are advected over $i$ cell boundaries with the transported volumes ${\rm Flux}_i$. 
Using this formula, we can now compute the kinetic energy effectively transported by the advection of mass and momentum.

In application to horizontal advection from one column to the other, we obtain for column 1 
\begin{equation}
\label{eq.kinenechange1}
 \widetilde{{\textstyle \frac{1}{2}} \rho u^2} {\rm V\_flux_\theta} \simeq
	 u_1                           \widetilde{\rho u} \; {\rm V\_flux_\theta} -
	{\textstyle \frac{1}{2}} u_1^2 \widetilde{\rho}   \; {\rm V\_flux_\theta}
\end{equation}
and column 2
\begin{equation}
\label{eq.kinenechange2}
 \widetilde{{\textstyle \frac{1}{2}} \rho u^2} {\rm V\_flux_\theta} \simeq
	 u_2                           \widetilde{\rho u} \; {\rm V\_flux_\theta} -
	{\textstyle \frac{1}{2}} u_2^2 \widetilde{\rho}   \; {\rm V\_flux_\theta} \: .
\end{equation}
Comparing these two lines, 
it becomes apparent that the advected kinetic energy $\widetilde{{\textstyle \frac{1}{2}} \rho u^2}$ 
is not identical in both columns: 
a certain flux of density $\widetilde{\rho}$ and momentum $\widetilde{\rho u}$ 
over the interface between the two columns 
induces a change in kinetic energy in column 1 as given by Eq.~\ref{eq.kinenechange1}, 
while column 2 experiences a change according to Eq.~\ref{eq.kinenechange2}. 
The advection process therefore creates an error in the kinetic energy balance, 
and consequently also in the conservation of total energy.

To allow for this deficit in the total energy balance, we compute the difference and place it 
as a source term into the equation of internal energy
\begin{equation}
   E_{\rm adv} = (u_1 - u_2) \, \widetilde{\rho u} \; {\rm V\_flux_\theta}  
               - \frac{1}{2} \left( u_1^2 - u_2^2 \right) \, \widetilde{\rho} \; {\rm S\_flux_\theta} \; .
\end{equation}
In the simplest case, horizontal transport uses donor cell advection or, more generally, a formalism as in Eq.~\ref{eq.rho-H-advec}.
Assuming that we compute $\widetilde{\rho}$ and $\widetilde{\rho u}$ analogously, i.e.\ with the same $\lambda$, 
we can further simplify
\begin{equation}
\label{eq.E_adv}
   E_{\rm adv} = (u_1 - u_2)^2 \, \rho^{\ast} \; {\textstyle \frac{1}{2}} \bigl| {\rm V\_flux_\theta} \bigr|
\end{equation}
where $\rho^{\ast}$ is given by
\begin{equation} 
\rho^{\ast} = \left\{ \begin{array}{cl}
	      (1-\lambda) \; \overline \rho_1 - \lambda \; \overline \rho_2 &\quad \mbox{for}\quad {\rm V\_flux_\theta} > 0 \\[2pt]
	      (1-\lambda) \; \overline \rho_2 - \lambda \; \overline \rho_1 &\quad \mbox{for}\quad {\rm V\_flux_\theta} < 0 \end{array} \right. 
\end{equation}
and $\overline \rho$ is the radially averaged density (see Sect.~\ref{sect.rho_avg}). 
For $\lambda = 0$ (donor cell), $\rho^{\ast}$ becomes the upstream value, 
in which case we could write $\rho^{\ast} = \widetilde{\overline \rho}$. 

From Eq.~\ref{eq.E_adv}, we have $E_{\rm adv} \ge 0$, 
i.e.\ $E_{\rm adv}$ always acts as a source term for the internal energy. 
$E_{\rm adv}$ therefore effectively describes the dissipation of kinetic 
energy in the course of advection from one column to another.
This dissipation increases with the radial velocity difference $\left| u_1 - u_2 \right|$ between the two columns. 
In a convection zone, the two columns hold opposing up- and downdraft motions and $\left| u_1 - u_2 \right|$ 
is quite large; $\left| u_1 - u_2 \right| \simeq 2 u_{\rm conv}$.
In these cases, the dissipation term becomes indispensable to the total energy balance;  
in the examples presented in Sect.~\ref{sect.results},  
it can account for more than 30\% of the energy throughput (i.e.\ luminosity).

Depending on the direction of the horizontal flow, the dissipated energy is deposited in the receiving column. 
In doing so, the contribution from Eq.~\ref{eq.E_adv} must be
divided radially to be consistent with the scalar discretization of the equation of internal energy.

\subsection{Radial distribution of grid points}

In the preceding paragraphs, we constructed the 
two columns discretization scheme with reference to 
a given radial distribution of the grid points $r_i$.
We now have to adopt a method to determine these grid point positions.

For obvious reasons, the convective fluid flow prohibits a Lagrangian grid customarily used in stellar models.
A spatially fixed Eulerian grid is also poorly suited to our needs for two reasons. 
First, advection alters the stellar structure. Starting from an initial, purely radiative model, 
the star shrinks significantly with the onset of advective transport.
Secondly, this scheme is intended to be used in computing stellar pulsations, 
i.e.\ to follow the convective circulating motion while the entire envelope 
moves in- and outward in the course of stellar pulsation.

To meet these requirements, the code uses an adaptive grid equation \citep{DD} 
that redistributes the grid points continuously
according to the evolving physical structures and therefore provides high 
resolution as needed, e.g.\ at photospheric gradients, 
while following radial movements of these features due to structural changes or stellar pulsation.
This adaptive grid equation is solved implicitly together with the physical equations.
Since the grid equation is an elliptic differential equation, 
this approach is only possible with an implicit solving method.

In the application to the two-column scheme, the same grid point distribution is used for both columns. 
Otherwise horizontal advection would become far more complicated and
 -- a serious issue for the implicit solving algorithm -- non-local with respect to the grid index $i$. 

Variables from both columns are used as `grid-weights', 
in particular the grid adapts according to gradients in density, internal energy, and $\nabla_{\rm ad}$ in each column.
The grid resolution is therefore increased in both columns in identical ways, even though, 
in general, only the physical structure in one of them actually demands this high grid resolution.

\section{Physical equations}
\label{sect.RHDeq-anal}

The physics within the two-column geometrical setup is computed from the 
equations of radiation hydrodynamics \citep[e.g.][]{MM}. 
The radiation field is thereby described using the 
first three gray moments of the intensity $J$, $\vec{H}$, and $\tens{K}$, 
which correspond to the radiative energy density, radiative flux, and radiative pressure, respectively.
An Eddington factor $f_{\rm edd}$ closes the moment equations.
Neglecting scattering, the source function of radiation, $S$, is given 
by the Stefan-Boltzmann law $S = \sigma/\pi T^4$, and 
$\kappa_{\rm R}$ and $\kappa_{\rm P}$ are the Rosseland and Planck mean opacities. 
The gas pressure $P$ and gas temperature $T$ are given by the equation of state. 
Self gravity is described by the gravitational potential $\phi$, which is assumed to be spherically symmetric, 
i.e.\ we do not allow for the (negligible) gravitational interaction between up- and downstreams. 
$G$ is Newton's gravitational constant.
Artificial viscosity, discussed in detail in Sect.~\ref{sect.visco}, 
enters in the form of the viscous pressure tensor $\tens{Q}$. \\

\noindent The system of analytical equations is given by: \\

\noindent {\it Equation of continuity}  
\begin{equation} 
\label{eq.cont_anal}
\frac{\partial}{\partial t} \rho
  + \vec \nabla \cdot ( \vec u \, \rho ) = 0 
\end{equation} 
{\it Equation of motion}
\begin{equation} 
\label{eq.mot_anal}
\frac{\partial}{\partial t} ( \rho \vec u )
   + \vec \nabla \cdot ( \vec u \, \rho \vec u ) 
   + \vec \nabla P 
   + \rho\vec{\nabla}\phi
   - \frac{4 \pi}{c} \kappa_{\rm R} \rho \vec{H}
   + \vec \nabla \cdot \tens{Q} = 0 
\end{equation} 
{\it Equation of internal energy}
\begin{equation}
\label{eq.ene_anal} 
\frac{\partial}{\partial t} ( \rho e )
  + \vec \nabla \cdot ( \vec u \, \rho e )
  + P \, \vec \nabla \cdot \vec u
  - 4 \pi \kappa_{\rm P} \rho ( J - S )
  + \tens{Q} : \vec \nabla \vec u    = 0
\end{equation} 
{\it Poisson equation}
\begin{equation}
\label{eq.mas_anal}
\Delta \phi = 4 \pi G \rho
\end{equation} 
{\it Radiation energy equation}
\begin{equation}
\label{eq.radene_anal} 
\frac{\partial}{\partial t} J
+ \vec \nabla \cdot ( \vec u \, J ) 
+ c \, \vec \nabla \cdot \vec H
+ \tens{K} : \vec \nabla \vec u
+ c\; \kappa_{\rm P} \rho ( J - S )  = 0
\end{equation} 
{\it Radiation flux equation}
\begin{equation}
\label{eq.radflx_anal} 
\frac{\partial}{\partial t} \vec H
+ \vec \nabla \cdot ( \vec u \, \vec H )
+ c \, \vec \nabla \cdot \tens{K} 
+ \vec H \cdot \vec \nabla \vec u 
+ c \; \kappa_{\rm R} \rho \vec H = 0
\end{equation}

\subsubsection*{Radiation equations for high optical depths}

The difference $(J-S)$ gradually vanishes with increasing optical depth, i.e.\ towards the interior of a star.

Therefore, the coupling term $(J-S)$ between radiative energy and gas energy becomes numerically unresolvable 
for high optical depths.
To derive still the correct contribution from this coupling for the equation of internal energy, 
the corresponding term $4\pi\kappa_{\rm P} \,  \rho \, (J-S)$
is expressed by the radiation energy equation 
and inserted into the equation of internal energy \citep{FeuchtingerptI}.
This corresponds to evaluating the sum `Equation of energy' + $\frac{4 \pi}{c}$ `Radiation energy equation', 
where terms with $(J-S)$ cancel out each other. 
The conversion factor $\frac{4 \pi}{c}$ relates the zeroth moment of the intensity $J$ to the radiation 
energy density
\begin{eqnarray}
\label{eq.ene_radene_anal}
& & \frac{\partial}{\partial t} ( \rho e + \textstyle{\frac{4 \pi}{c}} J )
  + \vec \nabla \cdot [ \vec u \, ( \rho e + \textstyle{\frac{4 \pi}{c}} J)] + \nonumber \\ & &
  + P \, \vec \nabla \cdot \vec u
  + \textstyle{\frac{4 \pi}{c}} \tens{K} : \vec \nabla \vec u 
  + 4 \pi \vec \nabla \cdot \vec H
  + \tens{Q} : \vec \nabla \vec u    = 0 \; .
\end{eqnarray}
Inwards of a predefined stellar depth, the equation of internal energy 
is substituted with this sum, i.e.\ Eq.~\ref{eq.ene_radene_anal} is solved instead of Eq.~\ref{eq.ene_anal}.

\section{Discrete set of equations}
\label{sect.RHDeq-discr}

After introducing the analytical form of the equations of radiation hydrodynamics, 
we now develop their discrete version.
Table~\ref{tab.primvar} summarizes the primary variables, 
the corresponding discrete equations, and the closures of the system.
As illustrated in Fig.~\ref{fig.2C-diskr-plus}, $r$ and $m$, 
as well as the `horizontal' variables $u_{\theta}$ and $H_{\theta}$, are integral quantities for both columns.
All other variables, $\rho$, $e$, $u$, $J$, and $H$ exist in duplicates, assigned individually to the two columns.

\begin{table*} 
\caption{Set of primary variables and the corresponding equations; 
for the discrete equations see also Table~\ref{tab.discSetofEq}.}
\label{tab.primvar}
\begin{center}
\begin{tabular}{l} 
\begin{tabular}{cll}
\hline
\noalign{\smallskip}
Variable                                        & Description & Equation \\
\noalign{\smallskip}
\hline
\noalign{\smallskip}
 $r_i$                                          & Radius
                                                & Adaptive grid equation \\
 $m_i$                                          & Integrated mass
                                                & Poisson equation, i.e.\ radial integration of mass \\
 $\rho_{1\,i}$, $\rho_{2\,i}$                   & Density 
                                                & Equation of continuity (in each column) \\ 
 $\overline\rho_{1\,i}$, $\overline\rho_{2\,i}$ & Averaged density
                                                & Radial averaging of $\rho$: Eq.~\ref{eq.rho1_avg}~\&~Eq.~\ref{eq.rho2_avg} \\
 $e_{1\,i}$, $e_{2\,i}$                         & Specific internal energy
                                                & Equation of energy (in each column) \\
 $u_{1\,i}$, $u_{2\,i}$                         & Radial velocity 
                                                & Equation of motion, radial component (in each column) \\
 $u_{\theta\,i}$                                & Horizontal velocity
                                                & Equation of motion, horizontal component \\
 $J_{1\,i}$, $J_{2\,i}$                         & $0^{\rm th}$ moment of radiation
                                                & Radiation energy equation (in each column) \\
 $H_{1\,i}$, $H_{2\,i}$                         & $1^{\rm st}$ moment of radiation, radial
                                                & Radiation flux Eq., radial component (in each column) \\
 $H_{\theta\,i}$                                & $1^{\rm st}$ moment of radiation, horizontal
                                                & Radiation flux Eq., horizontal component \\
\noalign{\smallskip}
\hline
\noalign{\smallskip}
\end{tabular} \\
\begin{tabular}{cl}
Closures:    
    & - tabulated equation of state (temperature, gas pressure), evaluated separately in each column: \\
    & $\qquad T_1 = T (\rho_1,e_1)$, $T_2 = T (\rho_2,e_2)$, $P_1 = P (\rho_1,e_1)$, $P_2 = P (\rho_2,e_2)$ \\
    & - tabulated opacities (Rosseland mean), evaluated separately in each column: \\
    & $\qquad \kappa_1 = \kappa (\rho_1,e_1)$, $\kappa_2 = \kappa (\rho_2,e_2)$ \\
    & - closure of radiation moments with an Eddington factor $f_{\rm edd} = \tens{K}/J = 1/3$ \\
\end{tabular} \\
\end{tabular}
\end{center}
\end{table*}

\subsection{Artificial viscosity}
\label{sect.visco}

In the continuum description of fluids, shock fronts -- and, in the present case, horizontal shear flows --
may become indefinitely sharp. 
In hydrodynamics codes, the smallest physical length scale is given by 
the mesh size of the numerical grid; 
on this length scale, numerical dissipation intrinsic to the spatial discretization becomes effective. 
In the present implementation,  
the adaptive grid continuously refines to resolve all gradients properly on the grid.
This reduces the intrinsic numerical dissipation and 
can thus lead to a runaway effect of successively steepening gradients and subsequent grid refinement.
It is therefore  necessary to include an artificial viscosity as a measure of broadening narrow physical 
features on a predefined length scale. 
Consequently, this also limits the maximum grid resolution to which the adaptive grid will be refined to.

In this way, artificial viscosity, by specifying the minimum length scale in the computation, 
plays a more important role than in usual Lagrangian or Eulerian hydrodynamics codes.

Due to the small overall dissipation of the numerical scheme, 
it is also sometimes necessary to include some extra viscosity 
to limit amplitudes and velocities, e.g.\ of stellar pulsations. 
However, in the results presented in Sect.~\ref{sect.results}, 
the influence of the artificial viscosity always remains negligible and is apparent only in 
a minor smoothing of velocity spikes.

For the artificial viscosity, the geometry-independent description provided by \citet{TW} was adopted.
In this description, modeled by analogy with the ordinary (molecular) fluid viscosity, the viscous pressure tensor reads
\begin{equation}
\label{eq.Q_anal}
\tens{Q} =  - \mu_{\rm Q} 
           \left( [\vec \nabla \vec u]_{\rm sym} - \mathbb{1} \, \frac{1}{3} \, \vec \nabla \cdot \vec u  \right)
\end{equation}
where the viscosity coefficient $\mu_{\rm Q}$ contains parameters 
for `linear' (pseudo-molecular) viscosity $q_{\rm lin}$ 
and `quadratic' viscosity $q_{\rm quad}$ 
(where `quadratic' refers to the quadratic dependency on the velocity field, 
which causes it to act in a way similar to a turbulent viscosity)
\begin{equation}
\label{eq.muQ}
\mu_{\rm Q} = q_{\rm lin}    l_{\rm visc}   \, \rho \, c_s 
            + q_{\rm quad}^2 l_{\rm visc}^2 \, \rho \, \max \left( -\vec \nabla \cdot \vec u,0 \right) \; .
\end{equation}
The use of the maximum implies that expanding flows are unaffected by viscosity.
The viscous length scale $l_{\rm visc}$ is set to the characteristic extension of the problem (and of the numerical grid), 
e.g.\ the radius in spherical geometry. $c_s$ is the local speed of sound.
For the symmetric velocity gradient, the notation $[\vec \nabla \vec u]_{\rm sym}$ was introduced. 
The symmetric description ensures that rotation, which does not affect the physical structure, 
remains unaffected by viscosity
\begin{equation}
\label{eq.nablaUsym_anal}
[\vec \nabla \vec u]_{\rm sym} = \frac{1}{2}\left( \vec \nabla \vec u + (\vec \nabla \vec u)^{\rm T} \right) \; .
\end{equation}
The contributions of artificial viscosity to the equations of motion and internal energy follow 
directly from the viscous pressure tensor. 
The viscous force is computed to be the divergence of the viscous pressure
\begin{equation}
\label{eq.forQ_anal}
\vec f_{\rm Q}  = \vec \nabla \cdot \tens{Q} \; .
\end{equation}
The viscous energy dissipation is obtained by contraction of the viscous pressure tensor 
with the gradient of the velocity field.
Since $\tens{Q}$ is symmetric, there is no difference between using the velocity gradient $\vec \nabla \vec u$ 
or the symmetric velocity gradient $[\vec \nabla \vec u]_{\rm sym}$ 
\begin{equation}
\label{eq.epsQ_anal}
\epsilon_{\rm Q} = \tens{Q} : \vec \nabla \vec u \; .
\end{equation}

To apply this recipe in the present case, Eqs.~\ref{eq.Q_anal}~--~\ref{eq.epsQ_anal} 
must be evaluated by assuming spherical geometry. 
Since the 2C-scheme describes all types of horizontal variability and 
dynamics with only one interface between the two radial columns, 
the $\theta$- and $\phi$-directions of spherical coordinates are not considered separately 
and we adopt the identities 
$u_\theta = u_\phi$ and $\frac{\partial}{\partial \theta} = \frac{\partial}{\partial \phi}$. 
To allow for that, all derivatives in the $\phi$-direction are assumed to be taken on the great circle, i.e.\ for $\theta = \pi/2$. 
Also note that $\tens{Q}$ is symmetric by definition and we therefore finally have four independent 
entries for the viscous pressure tensor in the two-column geometry: 
$Q_{rr}$, $Q_{r \theta}$, $Q_{\theta \theta}$, and $Q_{\theta \phi}$.

In principle, viscosity couples fluid flows in different coordinate directions. 
In the 2C-scheme, due to the combined discretization of $\theta$- and $\phi$-components, 
the corresponding terms in the spherical symmetric description become ambiguous in interpretation;
the two-column representation of the three-dimensional flow 
is too simplistic to enable a proper modelling of this effect. 
The viscous interaction between the two directions of fluid flow was therefore neglected by 
assuming $u_r = 0$ for the viscosity in the $\theta$-direction, and $u_\theta = 0$ in the radial direction.
For the viscosity in the radial direction, we then obtain
\begin{eqnarray}
\label{eq.2C-Qs-radial}
 Q_{rr}           &=& -\mu_Q \frac{2}{3}\left( \frac{\partial u_r}{\partial r} - \frac{u_r}{r} \right) \\
 Q_{r\theta}      &=& -\mu_Q \frac{1}{2} \frac{1}{r} \frac{\partial u_r}{\partial \theta}
\end{eqnarray}
\begin{equation}
\label{eq.2C-force-radial}
f_{{\rm Q}r} = \frac{3}{r} \frac{\partial}{\partial r^3} \left( r^3 \, Q_{rr} \right) 
             + \frac{2}{r} \frac{\partial Q_{r \theta}}{\partial \theta}
\end{equation}
\begin{equation}
\label{eq.2C-epsilon-radial}
\epsilon_{{\rm Q}r} = -\mu_Q \frac{2}{3}\left( \frac{\partial u_r}{\partial r} - \frac{u_r}{r} \right)^2
                      -\mu_Q \left( \frac{1}{r} \frac{\partial u_r}{\partial \theta} \right)^2 \; .
\end{equation}
For the viscosity in the $\theta$-direction, we arrive at
\begin{eqnarray}
\label{eq.2C-Qs-theta}
 Q_{r\theta}      &=& -\mu_Q \frac{1}{2}\left( \frac{\partial u_\theta}{\partial r} - \frac{u_\theta}{r} \right) \\
 Q_{\theta\theta} &=& -\mu_Q \frac{1}{3} \frac{1}{r} \frac{\partial u_\theta}{\partial \theta} 
\end{eqnarray}
\begin{equation}
\label{eq.2C-force-theta}
f_{{\rm Q}\theta} = 2\frac{3}{r} \frac{\partial}{\partial r^3} \left( r^3 \, Q_{r\theta} \right) 
                  + 2\frac{1}{r} \frac{\partial}{\partial \theta} \left( 4 Q_{\theta \theta} \right)
\end{equation}
\begin{equation}
\label{eq.2C-epsilon-theta}
\epsilon_{{\rm Q}\theta} = -\mu_Q \left( \frac{\partial u_\theta}{\partial r} - \frac{u_\theta}{r} \right)^2 
                           -\mu_Q 4\frac{2}{3}\left( \frac{1}{r} \frac{\partial u_\theta}{\partial \theta} \right)^2 \: .
\end{equation}
In Eq.\ref{eq.2C-force-theta}, an additional factor 2 was included for $f_{{\rm Q}\theta}$ because we are considering 
forces in both the $\theta$- and $\phi$-directions, even though they are combined in the discretization process.\\

In the discrete case, derivatives with respect to radius transform into differences between radial indices ($\Delta_r$), 
and derivatives in the $\theta$-direction are discretized using Eqs.~\ref{eq.derivtheta1}~\&~\ref{eq.derivtheta2}.

The various terms of the artificial viscosity (radial -- horizontal, shear -- non-shear)
include separate coefficients $\mu_{\rm Q}$ to allow for their individual adjustment.
Table~\ref{tab.viscopara} presents the $\mu_{\rm Q}$ coefficients with the parameters on which they depend.
The computation of the $\mu_{\rm Q}$ coefficients is similar to that described by Eq.~\ref{eq.muQ}, 
except for details related to the staggered-mesh location of the involved variables; 
the turbulent (`quadratic') viscosity parameter is only used for 
the radial, non-shear part ($\mu_{\rm Q1}$ and $\mu_{\rm Q2}$).
Where appropriate, the $\mu_{\rm Q}$'s are evaluated separately in each column,  
although the viscosity \emph{parameters} are the same in both columns.
In total, there are 5 viscosity parameters, although until now only three
(except for testing purposes) were actually used in the computations. 
The default values adopted in the examples presented in Sect.~\ref{sect.results}
are $q_{\rm lin} = 10^{-3}$, $q_{\rm quad} = 10^{-3}$, and $q_{\theta{\rm shear}} = 10^{-4}$.

\begin{table}
\caption{Compilation of viscosity coefficients and parameters.}
\label{tab.viscopara}
\begin{center}
\begin{tabular}{ccc}
\hline
\noalign{\smallskip}
Coefficient                  & Direction of action & Parameters   \\
\noalign{\smallskip}
\hline
\noalign{\smallskip}
$\mu_{\rm Q1}$               & radial              & $q_{\rm lin}$, $q_{\rm quad}$ \\
$\mu_{\rm Q2}$               & radial              & $q_{\rm lin}$, $q_{\rm quad}$ \\
$\mu_{\rm Q shear}$          & radial              & $q_{\rm shear}$ \\
\noalign{\medskip}
$\mu_{\rm Q \theta}$         & horizontal          & $q_{\theta{\rm lin}}$ \\
$\mu_{\rm Q \theta shear}$   & horizontal          & $q_{\theta{\rm shear}}$ \\
\noalign{\smallskip}
\hline
\end{tabular}
\end{center}
\end{table}

The discretization of the viscous force and energy dissipation uses 
the same discretization volumes as the corresponding equations, 
i.e.\ the equation of motion and equation of internal energy.
To emphasize the volume-integrated variables, the discrete forces and energies are denoted by capital letters; 
$F_{\rm Q} = \int f_{\rm Q} \, dV$ and $E_{\rm Q} = \int \epsilon_{\rm Q} \, dV$. 

For the viscous force in the radial direction, we obtain for column 1
\begin{eqnarray}
F_{\rm Q1} &=&
    - cf_1 \frac{8\pi}{3r} \Delta_r \left\{ \mu_{\rm Q1} \, \overline r^3 
           \left( \frac{\Delta_r u_1}{\Delta_r r} - \frac{\overline u_1}{\overline r} \right) \right\} \nonumber \\
& & + \mu_{\rm Q shear} \frac{N}{4 r^2} \left( u_1 - u_2 \right) \, \frac{1}{2}{\rm V\_vol} 
\end{eqnarray}
and for column 2
\begin{eqnarray}
F_{\rm Q2} &=&
    - cf_2 \frac{8\pi}{3r} \Delta_r \left\{ \mu_{\rm Q2} \, \overline r^3 
           \left( \frac{\Delta_r u_2}{\Delta_r r} - \frac{\overline u_1}{\overline r} \right) \right\}  \nonumber \\
& & - \mu_{\rm Q shear} \frac{N}{4 r^2} \left( u_1 - u_2 \right) \, \frac{1}{2}{\rm V\_vol} \; .
\end{eqnarray}
Note that both shear forces are discretized with $\frac{1}{2}{\rm V\_vol}$ instead of ${\rm V\_vol_1}$ and ${\rm V\_vol_2}$
to allow them to cancel out each other for the two columns.

The corresponding viscous energy dissipation reads
\begin{eqnarray}
E_{\rm Q1} &=&
    - \mu_{\rm Q1} \frac{2}{3}\left( \frac{\Delta_r u_1}{\Delta_r r} - \frac{\overline u_1}{\overline r} \right)^2 {\rm S\_vol_1} \nonumber \\
& & - \mu_{\rm Q shear} \frac{N}{8} \left( \frac{\overline u_1 - \overline u_2}{\overline r} \right)^2 {\rm S\_vol_1} \\
E_{\rm Q2} &=&
    - \mu_{\rm Q2} \frac{2}{3}\left( \frac{\Delta_r u_2}{\Delta_r r} - \frac{\overline u_1}{\overline r} \right)^2 {\rm S\_vol_2} \nonumber \\
& & - \mu_{\rm Q shear} \frac{N}{8} \left( \frac{\overline u_1 - \overline u_2}{\overline r} \right)^2 {\rm S\_vol_2} \; .
\end{eqnarray}
This formalism for the viscosity in the radial direction closely resembles -- except of course for the shear part -- the customary 
1D viscosity description given, e.g., by \citet{SaasFee} or \citet{FeuchtingerptI}.

In the horizontal direction, discretization yields a viscous force 
\begin{eqnarray}
F_{\rm Q \theta} &=&
   - \frac{2 \pi}{\overline r} \Delta_r \left\{ \mu_{\rm Q \theta shear} \, r^3 
              \left( \frac{\Delta_r u_\theta}{\Delta_r \overline r} - \frac{\overline u_\theta}{r} \right) \right\} \nonumber \\
& & + \mu_{\rm Q \theta} \frac{4 N}{3} \frac{u_\theta}{\overline r^2} \; {\rm H\_vol} \; ,
\end{eqnarray}
and a viscous energy dissipation  
\begin{eqnarray}
E_{\rm Q \theta 1} &=& 
    - \mu_{\rm Q \theta shear} \left( \frac{\Delta_r \overline u_\theta}{\Delta_r r} - \frac{u_\theta}{\overline r} \right)^2 {\rm S\_vol_1} \nonumber \\
& & - \mu_{\rm Q \theta} \frac{4 N}{3} \frac{u_\theta^2}{\overline r^2} \, {\rm S\_vol_1} \\
E_{\rm Q \theta 2} &=& 
    - \mu_{\rm Q \theta shear} \left( \frac{\Delta_r \overline u_\theta}{\Delta_r r} - \frac{u_\theta}{\overline r} \right)^2 {\rm S\_vol_2} \nonumber \\
& & - \mu_{\rm Q \theta} \frac{4 N}{3} \frac{u_\theta^2}{\overline r^2} \, {\rm S\_vol_2} \; .
\end{eqnarray}

\subsection{Radiative transport}
\label{sect.2C-radtrans}

In the moment description of radiation, 
the second moment of the intensity -- which corresponds to the radiation pressure -- 
is assumed to be of the following form
\begin{equation}
\label{eq.Ktensor}
\tens{K} = 
\left(
    \begin{array}{ccc}
       K_{rr}       &                  &              \\
                    & K_{\theta\theta} &              \\
                    &                  & K_{\phi\phi} \\
    \end{array}
\right) = 
\left(
    \begin{array}{ccc}
       K_{rr}       &                    &                    \\
                    & \frac{J-K_{rr}}{2} &                    \\
                    &                    & \frac{J-K_{rr}}{2} \\
    \end{array}
\right) 
\end{equation}
with the radial component given by a scalar Eddington factor
\begin{equation} 
K_{rr} = f_{\rm edd} J \; .
\end{equation}
The same Eddington factor is taken for both columns 
\begin{equation} 
K_{rr,1} = f_{\rm edd} J_1 \qquad
K_{rr,2} = f_{\rm edd} J_2 \; ,
\end{equation}
and, for the examples presented in this paper, it has been set 
to a constant value of $f_{\rm edd} = 1/3$ for simplicity.

The discrete equation of radiative flux in the horizontal direction (i.e.\ for $H_{\theta}$), 
does not use the full time-dependent equation Eq.~\ref{eq.radflx_anal}
but only its stationary part 
\begin{equation}
\label{eq.radflx-stationary_anal}
\vec \nabla \cdot \tens{K} + \kappa_{\rm R} \rho \vec H = 0 \; .
\end{equation}
In this way, we did not have to discretize the horizontal component of $\vec H \cdot \vec \nabla \vec u$,  
which is, in a similar way to artificial viscosity, 
ambiguous in interpretation in the context of the 2C-scheme. 
Considering the simplistic discretization of horizontal exchange between the two columns,
this stationary, diffusion-like description remains sufficient.

Note that the assumption for the radiative pressure in Eq.~\ref{eq.Ktensor} 
will in general not be consistent with the horizontal radiative flux 
computed from Eq.~\ref{eq.radflx-stationary_anal}.
However, a more consistent description is not reasonably possible 
given the limited resolution in horizontal direction of the 2C-scheme.
Adopting a more elaborate description would also require solving 
the detailed 2D radiative transport to obtain the required Eddington factors 
(e.g.\ $K_{rr} = f_{{\rm edd}rr} \,J$ and $K_{\theta\theta} = f_{{\rm edd}\theta\theta}\, J$).
And after all, there is no point in improving 
the radiative transport beyond the level of approximation of the hydrodynamics part.

\subsection{The discrete equations of radiation hydrodynamics}

Using the discretization scheme presented in Sect.~\ref{sect.2C-scheme} and the results from 
Sect.~\ref{sect.visco} and Sect.~\ref{sect.2C-radtrans}, 
we obtain the discrete version of Eqs.~\ref{eq.cont_anal}~--~\ref{eq.radflx_anal} \& Eq.~\ref{eq.radflx-stationary_anal}.
Table~\ref{tab.discSetofEq} provides the full discrete set of equations of radiation hydrodynamics. 
These physical equations are completed by the equations for the radially averaged densities, 
Eqs.~\ref{eq.rho1_avg}~\&~\ref{eq.rho2_avg}, and by the adaptive grid equation.

As an example of the discrete form of conservative equations  
and to illustrate the notation of the advective contributions, 
the discrete equations of continuity are given here for both columns
\begin{equation} 
\label{eq.cont1_discr}
\delta[\rho_1 \, {\rm S\_vol_1}] 
+ \Bigl[ \widetilde{\rho_1} {\rm S\_flux_1} \Bigr]_{i+1} 
- \Bigl[ \widetilde{\rho_1} {\rm S\_flux_1} \Bigr]_i 
+ \widetilde{\rho_1} {\rm S\_flux_\theta} = 0
\end{equation}
\begin{equation}
\label{eq.cont2_discr}
\delta[\rho_2 \, {\rm S\_vol_2}] 
+ \Bigl[ \widetilde{\rho_2} {\rm S\_flux_2} \Bigr]_{i+1} 
- \Bigl[ \widetilde{\rho_2} {\rm S\_flux_2} \Bigr]_i 
- \widetilde{\rho_2} {\rm S\_flux_\theta} = 0 \; .
\end{equation}
The notation $\delta[X]$ indicates a difference of $X$ between the new and old time level 
separated by the time step $\delta t$.
Advection represented by the terms with {\rm S\_flux},  occurs in the radial direction both at the radii $r_{i+1}$ and $r_i$ 
as well as in the horizontal direction over the interface between the two columns.
Note that the horizontal transport terms with ${\rm S\_flux_\theta}$ correspond to each other. 

In Table~\ref{tab.discSetofEq}, an abbreviated notation was adopted for 
the advective terms by summarizing all three contributions.
In this form, advective terms, e.g.\ those from Eq.~\ref{eq.cont1_discr}, are written as
\begin{equation} 
\sum \widetilde{\rho_1} \; {\rm S\_flux_{1,\theta}} \equiv \Bigl[ \widetilde{\rho_1} \; {\rm S\_flux_1} \Bigr]_{i+1} 
- \Bigl[ \widetilde{\rho_1} \; {\rm S\_flux_1} \Bigr]_i 
+ \widetilde{\rho_1} \; {\rm S\_flux_\theta} \: .
\end{equation}
Spatial differences are denoted as $\Delta_r$ and $\Delta_\theta$ in the radial and horizontal direction, respectively.
Averaged quantities -- where the precise definition depends on the context -- are written with overhead dashes.

\begin{table*} 
\caption{The discrete set of equations.}
\label{tab.discSetofEq}
{\it Equation of continuity} 
$$
\delta[\rho_1 \, {\rm S\_vol_1}] 
+ \Bigl[ \widetilde{\rho_1} \; {\rm S\_flux_1} \Bigr]_{i+1} 
- \Bigl[ \widetilde{\rho_1} \; {\rm S\_flux_1} \Bigr]_i 
+ \widetilde{\rho_1} \; {\rm S\_flux_\theta} = 0
$$
$$
\delta[\rho_2 \, {\rm S\_vol_2}] 
+ \Bigl[ \widetilde{\rho_2} \; {\rm S\_flux_2} \Bigr]_{i+1} 
- \Bigl[ \widetilde{\rho_2} \; {\rm S\_flux_2} \Bigr]_i 
- \widetilde{\rho_2} \; {\rm S\_flux_\theta} = 0
$$
{\it Integrated mass (Poisson equation)}
$$
\Delta_r m \, = \, \rho_1 \, {\rm S\_vol_1}  +  \rho_2 \, {\rm S\_vol_2} 
$$
{\it Equation of motion -- radial direction}
$$
\delta[\overline \rho_1 u_1 \, {\rm V\_vol_1}] 	
+ \sum \widetilde{(\overline \rho_1 u_1)} \; {\rm V\_flux_{1,\theta}} 
+ cf_1 4\pi \, r^2 \Delta_r (P_1) \, \delta t 
+ \frac{G m}{r^2} \, \overline \rho_1 \, {\rm V\_vol_1} \, \delta t 
- \frac{4\pi}{c}\overline{\kappa_1 \rho_1} \, H_1 \, {\rm V\_vol_1} \, \delta t
+ F_{\rm Q1} \, \delta t 
= 0 
$$
$$
\delta[\overline \rho_2 u_2 \, {\rm V\_vol_2}] 	
+ \sum \widetilde{(\overline \rho_2 u_2 )} \; {\rm V\_flux_{2,\theta}}
+ cf_2 4\pi \, r^2 \Delta_r (P_2) \, \delta t 
+ \frac{G m}{r^2} \, \overline \rho_2 \, {\rm V\_vol_2} \, \delta t
- \frac{4\pi}{c}\overline{\kappa_2 \rho_2} \, H_2 \, {\rm V\_vol_2} \, \delta t
+ F_{\rm Q2} \, \delta t
= 0 
$$
{\it Equation of motion -- horizontal direction}
$$
\delta[ \overline \rho_\theta u_\theta \, {\rm H\_vol}]
+ \sum \widetilde{(\overline \rho_\theta u_\theta)} \; {\rm H\_flux}
+ \frac{A_{\rm intf}}{2} \left( P_2 - P_1 \right) \, \delta t
- \frac{4\pi}{c} \left( cf_1 \kappa_1 \rho_1 + cf_2 \kappa_2 \rho_2 \right) \, H_\theta \, {\rm H\_vol} \, \delta t
- \frac{F_{\rm anhl}}{2}
+ F_{\rm Q \theta} \, \delta t 
= 0
$$
{\it Equation of energy}
$$
\delta[\rho_1 e_1 \, {\rm S\_vol_1}]
+ \sum \widetilde{(\rho_1  e_1)} \; {\rm S\_flux_{1,\theta}}
+ P_1 \, \left( cf_1 4\pi \Delta_r (r^2 u_1 ) + A_{\rm intf} u_\theta \right) \, \delta t
- 4\pi\kappa_1 \,  \rho_1 \, (J_1-S_1) \, {\rm S\_vol_1} \, \delta t 
- E_{\rm anhl} 
- E_{\rm adv}
+ E_{\rm Q1} \, \delta t 
+ E_{\rm Q \theta 1} \, \delta t
= 0 
$$
$$
\delta[\rho_2 e_2 \, {\rm S\_vol_2}]
+ \sum \widetilde{(\rho_2  e_2)} \; {\rm S\_flux_{2,\theta}}
+ P_2 \, \left( cf_2 4\pi \Delta_r (r^2 u_2) - A_{\rm intf} u_\theta \right) \, \delta t
- 4\pi\kappa_2 \,  \rho_2 \, (J_2-S_2) \, {\rm S\_vol_2} \, \delta t 
- E_{\rm anhl} 
- E_{\rm adv}
+ E_{\rm Q2} \, \delta t
+ E_{\rm Q \theta 2} \, \delta t
= 0 
$$
{\it Radiation energy equation}
$$
\delta[ J_1 \, {\rm S\_vol_1}]
+ \sum \widetilde{J_1} \; {\rm S\_flux_{1,\theta}}
+ c   \, \left( cf_1 4\pi \Delta_r (r^2 H_1) + A_{\rm intf} H_\theta \right) \, \delta t + $$ $$
+ cf_1 \, K_{rr,1} \, 4\pi \Delta_r (r^2 u_1) \, \delta t 
+ \left( J_1 - 3 K_{rr,1} \right) \frac{\overline u_1}{\overline r} \, {\rm S\_vol_1} \, \delta t
+ \frac{J_1 - K_{rr,1}}{2} A_{\rm intf} u_\theta \, \delta t
+ c \, \kappa_1 \, \rho_1 \, (J_1-S_1) \, {\rm S\_vol_1} \, \delta t = 0 
$$
$$
\delta[ J_2 \, {\rm S\_vol_2}]
+ \sum \widetilde{J_2} \; {\rm S\_flux_{2,\theta}}
+ c   \, \left( cf_2 4\pi \Delta_r (r^2 H_2) - A_{\rm intf} H_\theta \right) \, \delta t + $$ $$
+ cf_2 \, K_{rr,2} \, 4\pi \Delta_r (r^2 u_2) \, \delta t 
+ \left( J_2 - 3 K_{rr,2} \right) \frac{\overline u_2}{\overline r} \, {\rm S\_vol_2} \, \delta t
- \frac{J_2 - K_{rr,2}}{2} A_{\rm intf} u_\theta \, \delta t
+ c \, \kappa_2 \, \rho_2 \, (J_2-S_2) \, {\rm S\_vol_2} \, \delta t = 0 
$$
{\it Radiation flux equation -- radial direction}
$$
\delta [ H_1 \, {\rm V\_vol_1}]
+ \sum \widetilde{H_1} \; \; {\rm V\_flux_{1,\theta}}
+ cf_1 \, c \, 4\pi r^2 \Delta_r (K_{rr,1}) \, \delta t
+ c \, \frac{3\overline K_{rr,1} - \overline J_1}{r} \, {\rm V\_vol_1} \, \delta t
+ cf_1 \; 4\pi r^2 H_1 \, \Delta_r (\overline u_1) \, \delta t
+ c \, \overline{\kappa_1 \rho_1}  \, H_1 \, {\rm V\_vol_1} \, \delta t = 0 
$$
$$
\delta [ H_2 \, {\rm V\_vol_2}]
+ \sum \widetilde{H_2} \; \; {\rm V\_flux_{2,\theta}}
+ cf_2 \, c \, 4\pi r^2 \Delta_r (K_{rr,2}) \, \delta t
+ c \, \frac{3\overline K_{rr,2} - \overline J_2}{r} \, {\rm V\_vol_2} \, \delta t
+ cf_2 \; 4\pi r^2 H_2 \, \Delta_r (\overline u_2) \, \delta t
+ c \, \overline{\kappa_2 \rho_2}  \, H_2 \, {\rm V\_vol_2} \, \delta t = 0 
$$
{\it Radiation flux equation -- horizontal direction: stationary limit}
$$
\frac{A_{\rm intf}}{2} \Bigl[ \frac{J_2 - K_{rr,2}}{2} - \frac{J_1 - K_{rr,1}}{2} \Bigr]
+ \left( cf_1 \kappa_1 \rho_1 + cf_2 \kappa_2 \rho_2 \right) \, H_\theta \, {\rm H\_vol} = 0 
$$
\end{table*}

\subsection{The stencil}
\label{sect.stencil}

The discretization of the system of differential equations at the grid point $r_i$ 
also incorporates variables from adjacent grid locations. 
In the present case, dependencies are included up to a distance of two grid points.
Equations at the grid point $i$ may therefore include variables from $i-2$, $i-1$, $i$, $i+1$, $i+2$. 
Accordingly, this ensemble of five grid points is referred to as `5-point stencil'.

The shape of the stencil is correlated closely with the implicit solution method because it determines 
the structure of non-zero entries in the Jacobi matrix.
In the present implementation, the Jacobian is constructed `1D-style', 
i.e.\ all variables from \emph{both} columns (as assembled in Table~\ref{tab.primvar}) have only one running index, 
the radial grid point index $i$.
Alternatively, it would also be possible to use two running indices as in a 2D code,
the second having values of only 1 and 2 to differentiate between the two columns.
This type of indexing would 
assign fewer variables, only those from one column, to each pair of indices, 
but correspondingly also involve 
a larger $5\times2$ stencil and a significantly more complicated algorithm.
For 2D grids, this results in a Jacobian (composed of more numerous but smaller submatrices) 
that enables an increase of up to 50\% in the speed of the matrix inversion \citep{diss}.
However, in the present (extreme) case, where the grid has just two grid points in one direction, 
the inversion time is almost identical to the far simpler 1D-like discretization.

\subsection{Averaged density $\overline \rho$} 
\label{sect.rho_avg}

To develop an expression for the momentum in the radial direction for the equation of motion, 
the (scalar) densities must be averaged for the same (vector) localization of the velocities (see Fig~\ref{fig.2C-diskr-plus}).
As a second order advection scheme is used in the radial direction, 
the momentum -- and consequently the averaged density -- is required at 5 successive radius points.
Averaging for 5 successive points is not possible within the 5-point stencil, 
and therefore an additional variable, the radially averaged density $\overline \rho$, was introduced
\begin{eqnarray}
\label{eq.rho1_avg}
\overline\rho_1 &=& \frac{ \frac{1}{2} \left( {\rm S\_vol_1} \rho_1 \bigl|_i 
                                            + {\rm S\_vol_1} \rho_1 \bigr|_{i-1} \right) }{\rm V\_vol_1} \\
\label{eq.rho2_avg}
\overline\rho_2 &=& \frac{ \frac{1}{2} \left( {\rm S\_vol_2} \rho_2 \bigl|_i 
                                            + {\rm S\_vol_2} \rho_2 \bigr|_{i-1} \right) }{\rm V\_vol_2} \;.
\end{eqnarray}
These algebraic equations, Eq.~\ref{eq.rho1_avg}~\&~\ref{eq.rho2_avg}, are solved implicitly together with 
the system of discrete equations given in Table~\ref{tab.discSetofEq}.

\subsection{Boundary conditions}

Two successive ghost cells -- corresponding to the 5-point discretization -- constitute the boundary conditions 
in each column at both the inner and outer boundary. 

The inner boundary conditions are stated at a fixed inner radius of the computational domain and 
characterized by constant values for $\rho$, $e$, $m$, $J$, and $H$ (the same in both columns). 
The value of $H$ entering at the inner boundary corresponds to the luminosity of the modelled star; 
$m$ is the mass of the central core.
The radial velocities at the inner boundary are taken to be zero. 

For time-dependent computations of stellar pulsations -- the principle task to be solved by the code -- 
the entire envelope of the star must be considered. 
Since nuclear energy generation is not implemented in the code, 
it is impossible, however, to model the stellar core. 
Therefore, the inner boundary is usually placed as deep as possible,  
while remaining clear of the core region where nuclear burning might occur.
In the case of the Cepheid models presented in Sect.~\ref{sect.results}, the radius of the inner boundary 
was set to be 10\% of the photosphere radius. 

The outer boundary conditions are defined at the outermost grid point, which moves in a Lagrangian manner, i.e.\ there is no
fluid flow over the outer boundary to, or from, the exterior space. 
Accordingly, the radial velocities in the two columns are required to be identical at the outer boundary. 
A common equation of motion, formed as the sum by the individual equations of motions, 
determines the gas velocity at the outermost grid point
 -- and by means of the Lagrange condition -- the velocity of the grid point itself.
This setup has the advantage that the outer boundary of the grid can follow radius variations of the star 
e.g.\ due to stellar pulsations or structural resettling.
Obviously, there is no convective flux over the outer boundary.
The location of the outer boundary in relation 
to the mass structure is determined from the initial model 
and usually given by a predefined ratio (e.g.\ 1/100) between gas pressure 
at the outer boundary and the photospheric gas pressure.

For the physical conditions in exterior space, which affect the common equation of motion at the outermost grid point, 
$\frac{\partial \rho}{\partial r} = \frac{\partial e}{\partial r} = 0$ and $\tens{Q} = 0$ are assumed. 
These boundary conditions are, however, by no means unique and e.g.\ $\rho_{\rm ext} = {\rm const.}$ and $e_{\rm ext} = {\rm const.}$,  
or $\frac{\partial}{\partial r} \tens{Q} = 0$ would also be appropriate. 
When stellar pulsations are considered, these outer boundary conditions become more influential 
as they affect the wave reflection and dissipation properties. 

The boundary conditions for the radiation field assume free radiation at the outer boundary; 
$H$ is then computed to be $H = \mu \, \overline{J}$, where $\mu = \frac{1}{2}$ 
in the case of the Eddington approximation $f_{\rm edd} = \frac{1}{3}$, 
and $\overline{J}$ is a radially averaged value of $J$. 
This condition is evaluated individually for both columns, so that, in general, there 
will be a different radiative flux from each column.

\subsection{Temporal centering}

The system of equations of radiation hydrodynamics consists of parabolic differential equations.
Splitting them into a time derivative and spatial terms, they can be written in the form of
\begin{equation}
   \frac{\partial \vec X}{\partial t} = H(\vec X)
\end{equation}
where $H(\vec X)$ is a nonlinear spatial difference operator.
The time derivative is discretized to be $\delta[\vec X] / \delta t$
where $\delta t$ is the time step, and $\delta[\vec X]$ represents 
a difference in time between the new and old time level.
To achieve (almost) second order accuracy in time, the spatial terms 
must be evaluated centered in time, i.e.\ at a point in-between those two time levels in the temporal difference. 
This centering is completed in terms of variables, i.e.\ in the form of 
$H({\vec X}_{\rm cent.})$ with ${\vec X}_{\rm cent.} = 1/2 \left( \vec X^{\rm new} + \vec X^{\rm old} \right)$,   
as opposed to centering the operator ${H}_{\rm cent.}(\vec X)$. 
This centering of variables usually provides a higher temporal accuracy of the scheme \citep{Doetal}.
Based on the centered primary variables, successively all other required variables and expressions,  
such as cell volumes, advection fluxes, viscosity terms, opacities, and equation of state can be assembled.

\section{Method of solution}
\label{sect.method}

The system of nonlinear, discrete equations 
is solved time-dependently using an implicit Newton-Raphson iteration.
The implicit solution has the advantage of not being affected by the CFL time step limit \citep[after][]{CFL}
and also allows the inclusion of elliptical parts into the system of physical equations (Poisson and grid equation). 
The long time steps that are possible with the implicit scheme 
are particularly useful for the present problem of convective transport 
because they permit a rapid progression towards the stationary solution.

Each step in the Newton-Raphson iteration requires the inversion of the Jacobi matrix.
According to the system of 16 equations (Table~\ref{tab.primvar}), 
the Jacobian is composed of $16\times16$ submatrices, which form
a pentadiagonal structure of non-zeros reflecting the discretization with a 5-point stencil. 
The inversion of the Jacobi matrix uses the customary approach 
of a Newton-elimination of the two lower sub-diagonals,
followed by a back substitution of the resulting upper triangular matrix. 
Normalization of the Jacobian prior 
inversion, using the largest term in each discrete equation,  
significantly improves its numerical properties. 

The time step $\delta t$ used for advancing the system of physical equations
is regulated to maintain reasonable iteration numbers (usually between 2 and 4) 
and according to other requirements, 
e.g.\ limiting the relative changes in the primary variables per time step.
In case of divergences, the Newton-Raphson iteration is restarted with a reduced time step.
    
The crucial point about implicit methods is the computation of the derivatives required for the Jacobian. 
Derivation of the discrete physical equations (Table~\ref{tab.discSetofEq}) 
with respect to the primary variables leads to rather elaborate expressions. 
The implicit scheme is also very sensitive to errors and inaccuracies in these derivatives.
Computer algebra was therefore adopted to allow a fast and reliable computation of all required derivatives. 
These computer algebra scripts directly produce FORTRAN code that can be plugged-in into a source code. 
This feature proved to be very useful at the development stage 
because it facilitated numerous and quick tests of the discretization scheme.

The computing time for inversion of the Jacobi matrix scales with $np \times ng^3$ 
with $ng$ the number of equations (here 16) and $np$ the number of grid points (usually $np = 500$).
A current CPU at 3 GHz achieves about 10 iteration cycles (i.e.\ time steps) per second for this setup. 
Unfortunately, the inversion of the Jacobian does not parallelize efficiently. 
Nonetheless, the long time steps possible with the implicit solution method 
ensure that the 2C-scheme is much faster than `classical' explicit 2D or 3D hydrodynamics.

\section{Demonstrating example}
\label{sect.results}

\begin{figure}[thbp]
\begin{center}
\includegraphics[height=1.\linewidth, angle=-90]{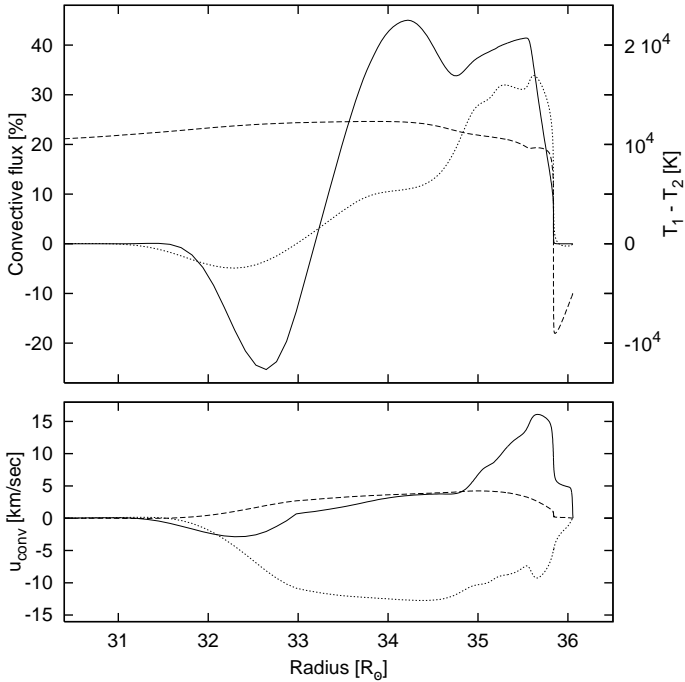}
\end{center}
\caption{Details of a Cepheid convection zone: 
The upper panel shows the convective transport 
in units of the total luminosity (solid line), the temperature difference between 
up- and downdrafts (dotted line) and the run of the entropy through the model (dashed line, without scale).
The convective velocities are given in the lower panel: updraft (dashed line), downdraft (dotted line) and horizontal (solid line).
A positive sign of the horizontal velocity corresponds to a flow from column 1 to column 2, i.e.\ from updraft to downdraft.
The figure focuses only on the outer convective region, the model actually extends down to about $3.6 \; R_{\odot}$.}
\label{fig.convzone} 
\end{figure}

According to the intention of applying this scheme in computations of Cepheid pulsations, 
a typical Cepheid with $T_{\rm eff}=5400\,K$, $L= 10^3\,L_{\odot}$, and $M=4.75\,M_{\odot}$
(which translates into $R_{\rm phot}=36.1\,R_{\odot}$ and $\log g = 2$)
was adopted for testing. 
This star, with a comparatively weak and shallow photospheric convection zone, 
has the advantage that it allows starting from a purely radiative initial model.  
For stars with fully convective envelopes, this is no longer possible because 
a purely radiative stratification would be too far off 
and therefore cause a violent collapse of the envelope with the onset of convection.
The inner boundary of the models was placed at 10\% of the photospheric radius (i.e.\ $\sim$$3.6 \; R_{\odot}$), 
although subsequent figures only indicate the outer convective region of interest.
The models consist of 500 radial grid points, the majority of which, due to the adaptive grid, cluster
around the photosphere and in the convective region.

To model a convection zone with wide up- and more narrowly confined downdrafts, 
updrafts are (arbitrarily) assigned to column 1 and the corresponding 
relative cross-section $cf_1$ is set to a value above $1/2$.
Accordingly, column 2 covers a smaller cross-section and contains the downdraft flows.
To ensure that convection finally occures in the intended sense of rotation, 
the initial model is perturbed with small radial velocities ($u \le 1\,\mathrm{m/sec}$) 
using the Schwarzschild convection criterion as a guide.

Starting the time-dependent simulation from that initial model, the convective velocity field develops rapidly and 
grows downwards from the photosphere. 
After a dynamic phase of growth that lasts about a thermal timescale 
of the relevant part of the envelope ($\sim$$10^7$ seconds), 
the convective velocities approach a stationary solution.
The time step then increases quickly and the computation is terminated at an age of $10^{12}$ seconds.
This evolution typically takes around a minute on a 3 GHz CPU and requires about 1000 time steps 
that increase in length 
during the computation from a few seconds at the start up to $10^{11}$ seconds for the stationary solution.

Figure~\ref{fig.convzone} shows the resulting convection zone using $N = 9951$ convective cells on a sphere, 
donor cell advection for the horizontal transport (Eq.~\ref{eq.rho-H-advec}, $\lambda = 0$), 
and a downstream cross-section of 20\% of the sphere ($cf_1 = 0.8$). 
As a useful guide, one can estimate the horizontal scale of 
photospheric convection (in the 2C-scheme, this corresponds to $D$, Eq.~\ref{eq.D}) to be about $10\, H_{p0}$ \citep{BF97}, 
where $H_{p0}$ is the characteristic photospheric pressure scale height, $H_{p0} = \mathcal{R}\,T_{\rm eff} / g$.
For the present example, a length scale of $20\, H_{p0}$ was adopted, which (evaluated at the photospheric radius)
translates into the aforementioned odd number of cells $N = 9951$.
This set of parameters, $D = 20 H_{p0}$, $cf_1 = 0.8$, $\lambda = 0$, serves subsequently 
as a reference for exploring the influence of the individual parameters.

The convection zone in Fig.~\ref{fig.convzone} includes the H/He~I as well as the He~II ionization zone. 
Both are apparent in the entropy profile given in the upper panel, the former causing the steep photospheric drop, 
the latter appearing as moderate gradient between 34 and 35 $R_{\odot}$. 
The continued gradient in-between those two ionization zones
(i.e.\ outwards about 35 $R_{\odot}$) is an effect of the convective transport and not present in purely radiative models.
The large temperature difference between up- and downdrafts in the outer part reflects a different radial 
position of the photosphere in the two columns. 

\begin{figure}[thbp]
\begin{center}
\includegraphics[height=.8\linewidth, angle=-90]{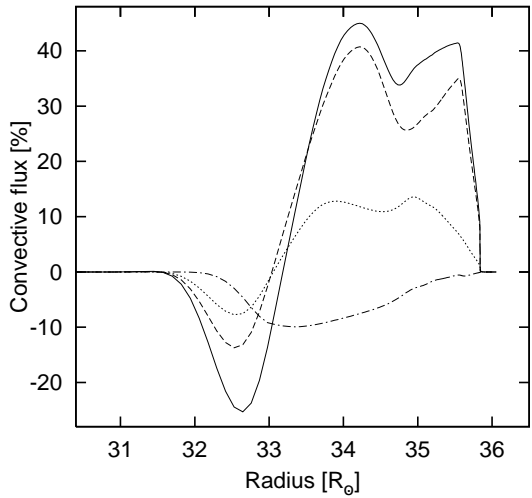}
\end{center}
\caption{Contributions to the convective energy flux: transport of internal energy by fluid motion (dashed line); 
kinetic energy flux (dash-dotted line); and flux due to work against gas, viscous, and radiative pressure (dotted line).
The radiative flux makes up for the difference between 
the sum of these species (solid line, commonly referred to as `convective energy flux') and 100\%.}
\label{fig.fluxcomp} 
\end{figure}

The convective velocities (Fig.~\ref{fig.convzone}, lower panel) show a comparatively slow updraft motion. 
In the thin outer regions -- around the photosphere --
the hot material loses energy by radiation and changes over to the downdraft column.
Due to the narrower downdrafts, the downward velocity is accordingly higher and
the large momentum in the downdraft motion produces a prominent inward overshoot. 
The mild entropy gradient in that part of the envelope 
also offers only little resistance to the downdrafts. 
The temperature difference is reversed in the overshoot; 
the downstream flow is now hotter than the `surroundings', and the convective flux has a negative sign.
Because of the contribution of the kinetic energy flux to the convective transport (see Fig.~\ref{fig.fluxcomp}), 
the temperature difference and the convective flux do not change their sign at exactly the same depth.
In the overshoot region, the material also returns to the updraft column, closing the circulating convective motion.

Figure~\ref{fig.fluxcomp} indicates the contributions to the convective flux: transport of internal energy, 
kinetic energy flux, and flux related to work against the total pressure 
(consisting of gas, viscous, and radiation pressure).
Even though viscous and radiative pressure have been included for completeness, 
the total pressure for this type of star is dominated largely by the gas pressure.
Radiation pressure accounts for up to about 15\% of the total pressure, viscous pressure for much less. 
Transport of potential energy is not evident in Fig.~\ref{fig.fluxcomp}, since the 
contributions from up- and downdrafts balance each other in the stationary case.
The flux of kinetic energy is entirely inward because of the 
narrower and more rapid downdrafts, which transport more kinetic energy than the updrafts.
This behavior of the kinetic energy flux 
is consistent with the results from multi-dimensional simulations of convection.

The ability of the code 
to reproduce the kinetic energy flux as well as the extended lower overshoot 
in qualitative agreement with multi-dimensional hydrodynamics computations 
\citep{Roxburgh1993,Muthsam1995,SFL2005} 
is an indication that the 2C-scheme succeeds in describing the essential physics of convective transport.

\subsection{Parameter studies}

\begin{figure}[thbp]
\begin{center}
\includegraphics[height=.8\linewidth, angle=-90]{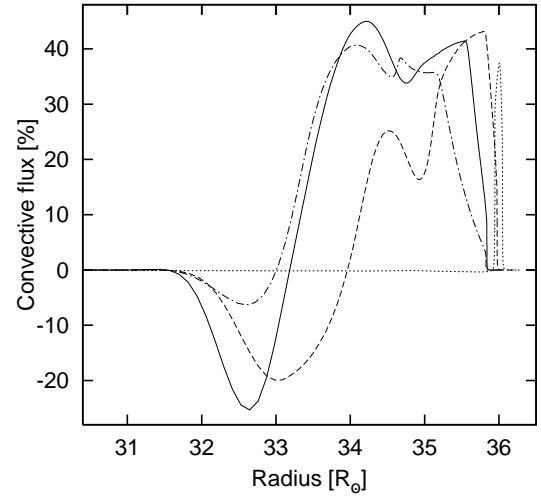}
\end{center}
\caption{Convective flux assuming the typical horizontal length scale $D$ to equal 
2 (dotted line), 10 (dashed line), 20 (solid line), and 80 (dash-dotted line) times the
characteristic photospheric pressure scale height $H_{p0}$.
}
\label{fig.nc} 
\end{figure}

The effect of the typical horizontal length scale $D$
 -- which translates into a certain number of convective cells $N$ on the sphere --
on the convective transport is shown in Fig.~\ref{fig.nc}.
For values of $D$ between $20 H_{p0}$ and $80 H_{p0}$, the convective flux is only slightly affected although 
the convective flux is somewhat lower for large convective cells, especially in the overshoot region. 
For even larger cells, it becomes increasingly more difficult 
for the convective circulation to 
bridge the growing distance between up- and downdrafts, and convection finally ceases. 
At the other extreme, convection also becomes less effective 
for convective cells smaller than $20 H_{p0}$. 
This seems reasonable as many thin downdrafts will dissolve rapidly, whereas a smaller number of more massive downdrafts
can retain their downward momentum much longer.
This causes the  H/He~I and He~II convection zones to separate, 
as is already apparent in Fig.~\ref{fig.nc} for the convective flux for $D = 10 H_{p0}$ (dashed line).
Ultimately, there remains only a narrow convective region related to H ionization 
as shown by the dotted line for $D = 2 H_{p0}$.

This transition from a large common convective region containing both the H/He~I and  He~II ionization zone 
to two decoupled convective shells 
also happens in a sequence of models when changing to `less-convective' stellar parameters (e.g.\ higher effective temperature).
Usually -- at least for the investigated Cepheid-like stars -- the inner He~II convection 
carries only marginal flux, although showing convective velocities of a several $\mathrm{km/sec}$.
These decoupled convection zones found for hotter Cepheids are similar 
to those obtained for A-type stars \citep{KuMo2002,SFL2005}.
The difference in effective temperature of about 2000~K between hot Cepheids and cool A-type stars 
appears to be largely compensated by the higher surface gravity of the A-type stars.

\begin{figure}[thbp]
\begin{center}
\includegraphics[height=.8\linewidth, angle=-90]{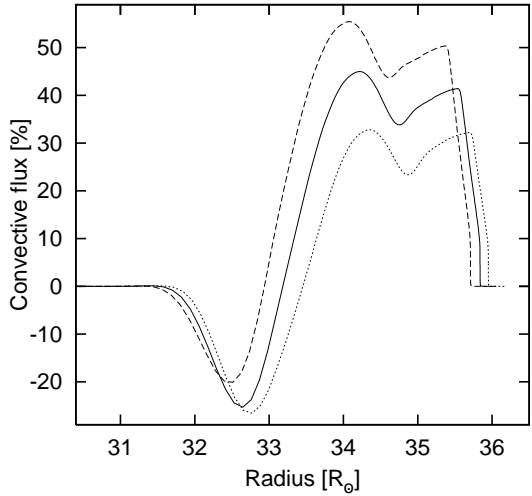}
\end{center}
\caption{Influence of the ratio of cross-section between up- and downdraft. 
The plot gives the convective fluxes of convection zones 
where the downdrafts take 25\% (dashed line), 20\% (solid line), and 15\% (dotted line) of the sphere 
(i.e.\  $cf_1$ is 0.75, 0.8, and 0.85).}
\label{fig.cf1} 
\end{figure}

In Fig.~\ref{fig.cf1}, the effect of $cf_1$ is studied by showing convection zones 
with downdrafts taking 25\%, 20\%, and 15\% of the sphere.
Note that more narrow downdrafts lead, due to correspondingly more rapid downdraft motion,  
to a more pronounced overshoot despite a reduced overall convective efficiency. 
Concerning the efficiency of convection, a 50/50 ratio of up- and downstream cross-section would obviously be the optimum, 
but that is probably not a realistic scenario for photospheric convection in real stars.
In contrast to the horizontal length scale $D$, for which the hydrostatic pressure scale height $H_{p0}$ 
provides good indications of a reasonable parameter range, 
the proper value of $cf_1$ is more difficult to estimate and requires further investigation.
The granulation pattern of the Sun as well as multi-dimensional hydrodynamics computations 
of other stars \citep[e.g.][]{BF96,SFL2005} clearly suggest rather narrow downdrafts.

\begin{figure}[thbp]
\begin{center}
\includegraphics[height=.8\linewidth, angle=-90]{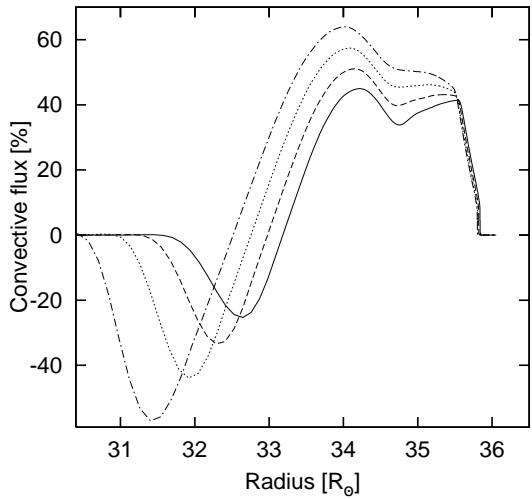}
\end{center}
\caption{Effect of the parameter $\lambda$ for the horizontal advection (see Eq.~\ref{eq.rho-H-advec}). 
The figure shows the relative convective flux for $\lambda$ equalling 
$0$ (i.e.\ donor cell, solid line), $0.05$ (dashed line), $0.1$ (dotted line), and $0.15$ (dash-dotted line).}
\label{fig.advH} 
\end{figure}
Figure~\ref{fig.advH} illustrates the effect of the horizontal advection scheme quantified by the
parameter $\lambda$ as described in Sect.~\ref{sect.horizAdv}.
As already argued there, one should keep $\lambda$ well below $0.5$.
Increasing $\lambda$ from $0$ to $0.5$ changes the horizontal advection from donor cell to centered values, 
which successively reduces the dissipation of radial momentum due to horizontal exchange. 
Consequently, the downdraft moment is retained longer when the convective circulation makes its turnaround 
in the lower overshoot region. 
The effect of increasing $\lambda$ is therefore basically a deeper overshoot as well as a higher 
overall convective efficiency because of the reduced dissipation.

Summarizing the discussion of the parameter influence 
and considering qualities such as convective flux, depth of the convective region, and amount of overshoot, 
it appears that, although there is some variability in the results, the basic type of solution is quite robust.
It is possible to suppress convection by choosing extreme parameters,
some combinations of parameters may also cause numerical problems; 
none of the test computations, however, produced a convective region qualitatively different 
from those shown in Figs.~\ref{fig.nc}~--~\ref{fig.advH}. 

Even though the adopted parameters are up to now little more than an educated guess
and still require verification by comparison with observations or more elaborate numerical simulations, 
the `reasonable parameter range' suggested here is probably quite reliable.

\subsection{Accuracy of the discretization}

\begin{figure}[thbp]
\begin{center}
\includegraphics[height=.8\linewidth, angle=-90]{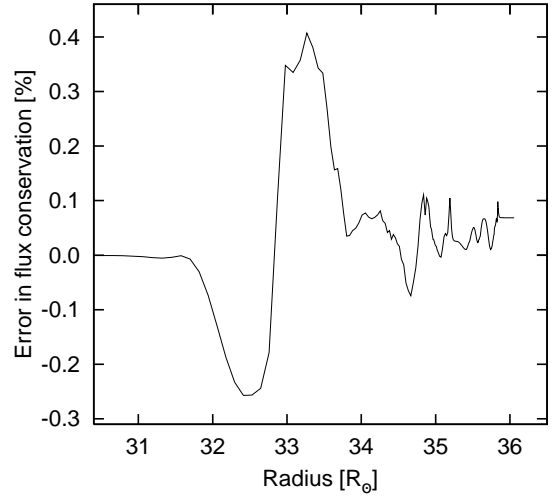}
\end{center}
\caption{Accuracy of flux conservation throughout the convection zone presented in Fig.~\ref{fig.convzone}.
For stationary solutions, deviations from a constant energy flux correspond to errors in the total energy conservation.
Inwards of the plotted region, the flux is entirely transported by radiation and hence no errors occur.
The same is true for the outermost part of the model above the convective region.}
\label{fig.fluxconst} 
\end{figure}
According to the nature of convection, a considerable part of the luminosity 
is converted from radiation to internal and kinetic energy, transported upwards through the convection zone,  
where radiation once again takes over.
This conversion of energy causes errors in the total energy balance, which are eventually evident in the total 
energy flux (i.e.\ luminosity) for stationary solutions.
In the present type of discretization, the \emph{total} energy is not treated conservatively 
but is composed of several species (internal, kinetic, potential, and radiative energy); 
the total energy conservation is therefore a 
good measure of the discretization accuracy and a possible way of testing physical soundness.

The deviations from constant total energy flux are shown in Fig.~\ref{fig.fluxconst} for the standard-parameter convection zone.
Without the correction term accounting for the momentum dissipated by horizontal advection given by Eq.~\ref{eq.E_adv}, 
the discrepancies would become larger than 30\%.

\section{Conclusions}
\label{sect.concl}

The scheme proposed in this paper has a number of advantages:
\begin{itemize}
\item It is a non-local description of convection 
      and therefore provides a consistent computation of the depth of the convective region
      including convective overshoot.
\item The convective flux is computed directly from hydrodynamics and not from a heuristic, parametrized model.
\item The two-column convection has stationary solutions and in principle allows arbitrarily large time steps.
      This implies that it is suitable for application to problems involving long time series, 
      such as stellar pulsations or stellar evolution.
\item The 2C-model is much faster than multi-dimensional hydrodynamics computations; 
      stationary solutions can be obtained within minutes.
\item Radiative transport is an intrinsic part of the scheme, i.e.\ no hydrodynamical model with plugged-in radiation effects.
\item The main parameters of the scheme have a straightforward geometrical meaning 
      that also provides indications of reasonable values for these parameters.
\end{itemize}
However, there are also shortcomings to be considered:
\begin{itemize}
\item The 2C-scheme is basically still a parameter-dependent model. 
      These parameters require proper adjustment.
\item Horizontal advection and radiative transport are poorly represented 
      because of the very coarse `two-cell' spatial resolution in the horizontal direction.
\item The 2C-model uses a simplistic description of the full spectrum of vertical and horizontal convective motion, 
      which ignores turbulence effects and limits the investigation of more subtle features of convection. 
\end{itemize}
   
In its present form, the two-column scheme provides a simple, yet physically sound and consistent, non-local, 
radiation-hydrodynamics description of the convective circulation.
The model still contains free parameters, 
but their geometrical interpretation provides at least reasonable indications of proper values, 
and  they do not change the results by magnitudes.
For applications in which more detail and higher certainty is required, more elaborate methods, 
such as multi-dimensional hydrodynamics or turbulence models, remain the most appropriate alternative.


\begin{acknowledgements}
The author would like to thank B. Freytag for numerous helpful and inspiring discussions 
and for sharing his insights into the nature of convection.
This work was funded by Agence Nationale de la Recherche
under the ANR project number NT05-3~42319.
\end{acknowledgements}




\begin{thebibliography}{}


\bibitem[Abdella \& McFarlane(1997)]{AMF1997}
   Abdella, K.; McFarlane, N. : 1997, J.~Atmos.~Sci., 54, 1850
\bibitem[Asplund et al.(2000)]{Asplund2000} 
   Asplund, M.; Nordlund, \AA.; Trampedach, R.; Allende Prieto, C.; Stein, R. F.: 2000, A\&A, 359, 729
\bibitem[B\"ohm-Vitense(1958)]{MLT} 
   B\"ohm-Vitense, E.: 1958, Zs.Ap., 46, 108
\bibitem[Bono \& Stellingwerf(1994)]{BS1994}
   Bono, G.; Stellingwerf, R.F.: 1994, ApJS, 93 233
\bibitem[Canuto \& Mazzitelli(1991)]{CM91}
   Canuto, V.M.; Mazzitelli, I.: 1991, ApJ, 370, 295
\bibitem[Canuto(1992)]{canuto1992}
   Canuto, V.M.: 1992, ApJ, 392, 218
\bibitem[Canuto(1993)]{canuto1993}
   Canuto, V.M.: 1993, ApJ, 416, 331
\bibitem[Canuto(1996)]{C96}
   Canuto, V.M.: 1996, ApJ, 467, 385
\bibitem[Canuto et al.(1996)]{CGM96}
   Canuto, V.M.; Goldman, I.; Mazzitelli, I.: 1996, ApJ, 473, 550
\bibitem[Canuto(1997)]{canuto1997}
   Canuto, V.M.: 1997, ApJ, 482, 827
\bibitem[Canuto et al.(2007)]{CCH2007}
   Canuto, V.M.; Cheng, Y.; Howard, A.M.: 2007, Ocean~Modell., 16, 28
\bibitem[Chatfield \& Brost(1987)]{CB1987}
   Chatfield, R.B.; Brost, R.A.: 1987, J.~Geophys.~Res., 92, 13263
\bibitem[Courant, Friedrichs \& Lewy(1928)]{CFL}
   Courant, R.; Friedrichs, K.; Lewy, H.: 1928, Math. Ann., 100, 32
\bibitem[Cox \& Giuli(1968)]{CoxandGiuli}
   Cox, J.P.; Giuli, R.T.: 1968, Principles of Stellar Structure, Vol. I, Gordon and Breach, New York
\bibitem[Dorfi(1998)]{SaasFee}
   Dorfi, E.A.: 1998, in Computational Methods for Astrophysical Fluid Flow, Saas-Fee Advanced Course 27, Springer, Berlin, p.~263
\bibitem[Dorfi \& Drury(1987)]{DD}
   Dorfi, E.A.; Drury, L.O'C.: 1987, J.~Comp.~Phys., 69, 175
\bibitem[Dorfi et al.(2006)]{Doetal}
   Dorfi, E.A.; Pikall, H.; St\"okl, A.; Gautschy, A.: 2006, Comp.~Phys.~Comm., 174, 771
\bibitem[Feuchtinger(1999a)]{FeuchtingerptI}
   Feuchtinger, M.U.: 1999a, A\&AS, 136, 217
\bibitem[Feuchtinger(1999b)]{FeuchtingerptII}
   Feuchtinger, M.U.: 1999b, A\&AS, 351, 103
\bibitem[Freytag et al.(1996)]{BF96}
   Freytag, B.; Ludwig, H.-G.; Steffen, M.: 1996, A\&A, 313, 497
\bibitem[Freytag et al.(1997)]{BF97}
   Freytag, B.; Holweger, H.; Steffen, M.; Ludwig, H.-G.: 1997, in Science with the VLT Interferometer, ed. F. Paresce, Springer, Berlin, p.~316
\bibitem[Gehmeyr \& Winkler(1992)]{GM92}
   Gehmeyr, M.; Winkler, K.-H.A.: 1992, A\&A, 253, 92
\bibitem[Gryanik \& Hartmann(2002)]{GH2002}
   Gryanik, V.M.; Hartmann, J.: 2002, J.~Atmos.~Sci., 59, 2729
\bibitem[Gryanik et al.(2005)]{GHRS2005}
   Gryanik, V.M.; Hartmann, J.; Raasch, S.; Schr\"o ter, M.: 2005, J.~Atmos.~Sci., 62, 2632
\bibitem[Koll\' ath et al.(2002)]{KBSC2002}
   Koll\' ath, Z.; Buchler, J.R.; Szab\' o, R.; Csubry, Z.: 2002, A\&A, 385, 932
\bibitem[Kupka(1999)]{Kupka99}
   Kupka, F.: 1999, ApJ, 526, L45
\bibitem[Kupka \& Montgomery(2002)]{KuMo2002}
   Kupka, F.; Montgomery, M.H.: 2002, MNRAS, 330, L6
\bibitem[Kuhfu\ss(1986)]{kuhfuss}
   Kuhfu\ss, R.: 1986, A\&A, 160, 116
\bibitem[Lappen \& Randall(2001)]{LR2001}
   Lappen, C.-L.; Randall, D.A.: 2001, J.~Atmos.~Sci., 58, 2021
\bibitem[Lesaffre et al.(2005)]{lesaffre2005}
   Lesaffre, P.; Podsiadlowski, Ph.; Tout, C.A.: 2005, MNRAS, 356, 131 
\bibitem[Margrave \& Swihart(1969)]{MG1969}
   Margrave, T.E.; Swihart, T.L.: 1969, Solar~Phys., 6, 12
\bibitem[Mihalas \& Mihalas(1984)]{MM}
   Mihalas, D.; Mihalas, B.W.: 1984, Foundations of Radiation Hydrodynamics, Oxford University Press, New York
\bibitem[Montgomery \& Kupka(2002)]{MoKu2004}
   Montgomery, M.H.; Kupka, F.: 2004, MNRAS, 350, 267
\bibitem[Morton et al.(1956)]{MTT1956}
   Morton, B.R.; Taylor, G.I.; Turner, J.S.: 1956, Proc.~Roy.~Soc.~London, 234, 1
\bibitem[Muthsam et al.(1995)]{Muthsam1995}
   Muthsam, H.J.; G\"ob W.; Kupka, F.: Liebich, W.; Z\"ochling, J.: 1995, A\&A, 293, 127
\bibitem[Nordlund(1976)]{nordlund76}
   Nordlund, \AA.: 1976, A\&A, 50, 23
\bibitem[Nordlund et al.(1997)]{nordlundetal1997}
   Nordlund, \AA.; Spruit, H.C.; Ludwig, H.-G.; Trampedach, R.: 1997, A\&A, 328, 229
\bibitem[Prandtl(1925)]{Prandtl1925}
   Prandtl, L.: 1925, Z.~angew.~Math.~Mech., 5, 136
\bibitem[Randall et al.(1992)]{RSM1992}
   Randall, D.A.; Shao, Q., Moeng, C.-H.:  1992, J.~Atmos.~Sci., 49, 1903
\bibitem[Roxburgh \& Simmons(1993)]{Roxburgh1993}
   Roxburgh, I.W.; Simmons, J.: 1993, A\&A, 277, 93
\bibitem[Steffen et al.(2005)]{SFL2005}
   Steffen, M.; Freytag, B.; Ludwig, H.-G.: 2005, in Proc. 13th Cool Stars Workshop, eds. F. Favata et al., ESA SP-560, p.~985
\bibitem[Stein \& Nordlund(1998)]{SN98}
   Stein, R. F.; Nordlund, \AA.: 1998, ApJ, 499, 914
\bibitem[Stellingwerf(1982)]{stell}
   Stellingwerf, R.F.: 1982, ApJ, 262, 330
\bibitem[St\"okl(2006)]{diss}
   St\"okl, A.: 2006, PhD thesis, University of Vienna
\bibitem[Telford(1970)]{Telford1970}
   Telford, J.W.: 1970, J.~Atmos.~Sci., 27, 347
\bibitem[Tscharnuter \& Winkler(1979)]{TW}
   Tscharnuter, W.M.; Winkler, K.-H.A..: 1979, Comp.~Phys.~Comm., 18, 171
\bibitem[van~Leer(1974)]{vL74}
   van~Leer, B.: 1974, J.~Comp.~Phys., 14, 361
\bibitem[van~Leer(1977)]{vL77}
   van~Leer, B.: 1977, J.~Comp.~Phys., 23, 276
\bibitem[Wang \& Albrecht(1986)]{WA1986}
   Wang, S; Albrecht, B.A.: 1986, J.~Atmos.~Sci., 43, 2374
\bibitem[Wedemeyer et al.(2004)]{Wedemeyer2004}
   Wedemeyer, S.; Freytag, B.; Steffen, M.; Ludwig, H.-G.; Holweger, H.: 2004, A\&A, 414, 1121
\bibitem[Xiong(1989)]{Xiong1989}
   Xiong, D.R.: 1989, A\&A, 209, 126
\bibitem[Xiong et al.(1997)]{Xiongetal1997}
   Xiong, D.R.; Cheng, Q.L.; Deng, L.: 1997, ApJS, 108, 529
\bibitem[Zilitinkevich et al.(1999)]{ZGLM1999}
   Zilitinkevich, S.S.; Gryanik, V.M.; Lykossov, V.N.; Mironov, D.V.: 1999, J.~Atmos.~Sci., 56, 3463

\end{thebibliography}
\end{document}